\documentclass[english]{article}
\usepackage[T1]{fontenc}
\usepackage[utf8]{inputenc}
\usepackage{geometry}
\usepackage{verbatim}
\usepackage{amsmath}
\usepackage{amssymb}
\usepackage{graphicx}
\usepackage{setspace}
\usepackage{authblk}
\usepackage{mathtools}
\usepackage[dvipsnames]{xcolor}
\usepackage{siunitx}
\usepackage{float}
\usepackage{subfig}
\usepackage[normalem]{ulem}
\makeatletter

\date{}

\usepackage{hyperref}
\usepackage{siunitx}
\usepackage{bm}
\usepackage{tabularx}
\usepackage{cancel}
\usepackage{siunitx}
\usepackage[numbers]{natbib}

\@ifundefined{showcaptionsetup}{}{%
 \PassOptionsToPackage{caption=false}{subfig}}
\usepackage{subfig}
\makeatother

\usepackage{babel}

\begin{document}
\title{A Micromechanical Model for Light-interactive Molecular Crystals}
\author{\large{Devesh Tiwari}}
\author{\large{Ananya Renuka Balakrishna}\thanks{Corresponding author: ananyarb@ucsb.edu}}
\affil{\textit{Materials Department, University of California, Santa Barbara, CA 93106}}


\maketitle
\begin{abstract}

Molecular crystals respond to a light stimulus by bending, twisting, rolling, jumping, or other kinematic behaviors. These behaviors are known to be affected by, among others, the intensity of the incident light, the aspect ratios of crystal geometries, and the volume changes accompanying phase transformation. While these factors, individually, explain the increase in internal energy of the system and its subsequent minimization through macroscopic deformation, they do not fully explain the diversity of deformations observed in molecular crystals. Here, we propose a micromechanical model based on the Cauchy-Born rule and photoreaction theory to predict the macroscopic response in molecular crystals. By accounting for lattice geometry changes and microstructural patterns that emerge during phase transformation, we predict a range of deformations in a representative molecular crystal (salicylideneamine). Doing so, we find that the interplay between photoexcited states and the energy minimization pathways, across a multi-well energy landscape, is crucial to the bending and twisting deformations. We use our model to analyze the role of particle geometries and the intensity of incident light on macroscopic deformation, and identify geometric regimes for shearing and twisting deformations in salicylideneamine crystals. Our micromechanical model is general and can be adapted to predict photomechanical deformation in other molecular crystals undergoing a solid-to-solid phase change and has potential as a computational design tool to engineer reversible and controllable actuation in molecular crystals.

\end{abstract}

\section{Introduction}

\begin{figure}
    \centering
    \includegraphics[width=0.6\textwidth]{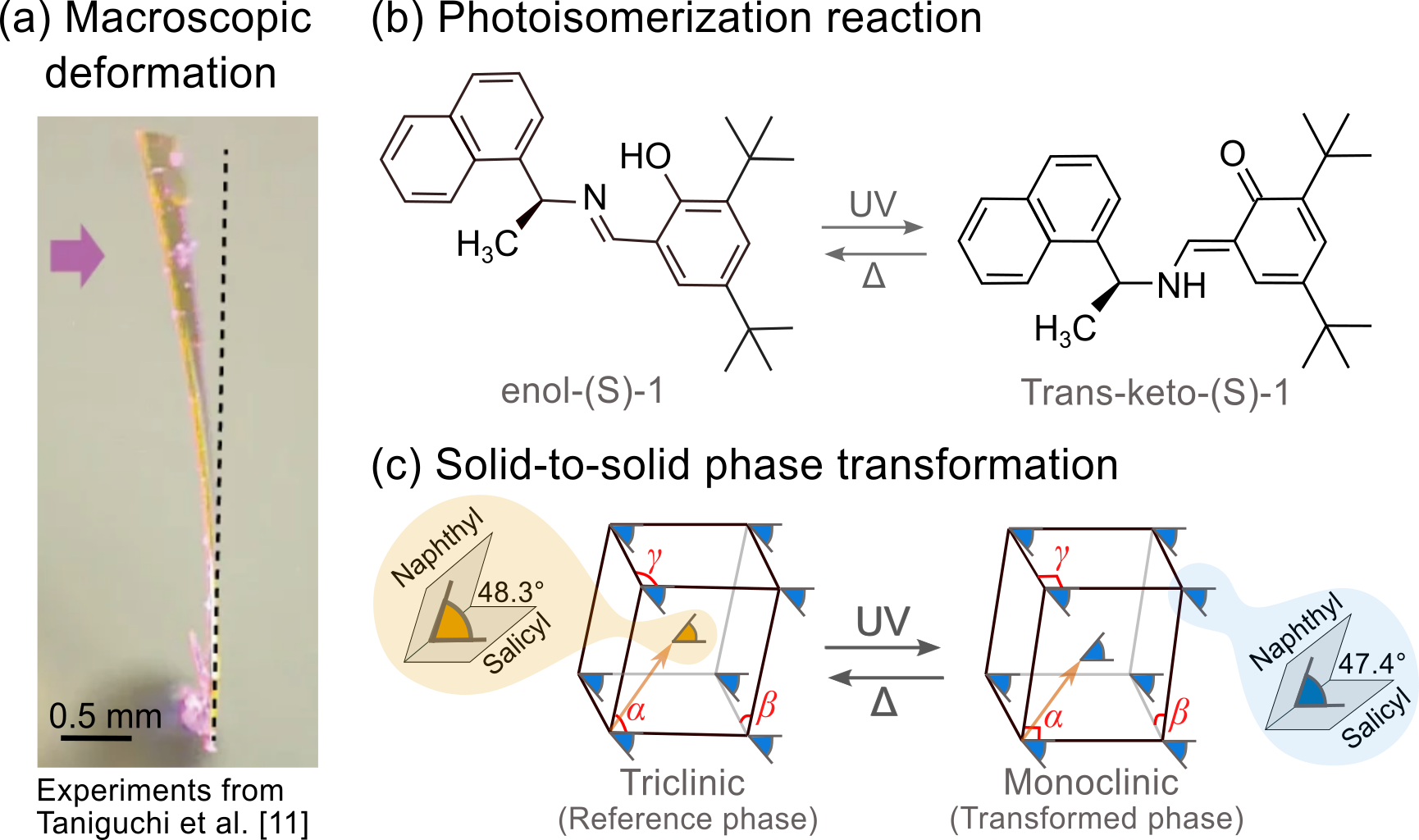}
    \caption{(a) Molecular crystals, such as the salicylideneamine, undergo a bending and twisting deformation on exposure to a light stimulus \cite{taniguchi2019photo}. Reprinted from Ref.~\cite{taniguchi2019photo} under a \href{https://creativecommons.org/licenses/by/4.0/}{Creative Commons Attribution 4.0 International License}.  (b) Illuminating a salicylideneamine crystal with UV light ($\lambda = 365\mathrm{nm}$) induces an enol-to-keto photoisomerization reaction of individual molecules. The photoexcited keto-molecules trigger a solid-to-solid phase transformation of the molecular crystal. (c) A schematic illustration of unit cells of the salicylideneamine crystal undergoing a displacive-type of phase transformation. In the reference phase, individual salicylideneamine molecules are arranged in triclinic symmetry, which, on transformation, deforms displacively to monoclinic symmetry. In addition to the change in unit cell geometry, phase transformations in the salicylideneamine crystals are accompanied by a conformational change of select molecules. We denote these conformational changes (i.e., a change in dihedral angles $48.3^\circ \to 47.4^\circ$ between the salicyl and naphthyl planes) using the yellow and blue angle notations, respectively \cite{taniguchi2019photo}.}
    \label{fig:isomerization}
\end{figure}

Molecular crystals that respond to a light stimulus represent an exciting class of photomechanical materials, with engineering applications spanning remote actuation, flexible electronics, and microdevices \cite{naumov2020rise}. These materials consist of discrete molecules held together by intermolecular interactions and display a long-range order that is intermediate to inorganic crystals and polymers (see {Fig.~\ref{fig:crystalline-order}}). On exposure to light, individual molecules undergo chemical reactions (e.g., enol-keto or cis-trans isomerization \cite{reviewisoPhase}, see Fig.~\ref{fig:isomerization}), which cooperatively induce a solid-to-solid transformation of the unit cells. These structural changes are abrupt and play an important role in shaping the photomechanical response of molecular crystals, ranging from simple stretching to more complex deformation such as coiling and twisting \cite{naumovReview, crystaltwist, crystalcurl}. In extreme cases, these crystals exhibit kinematic effects such as jumping to distances $10^5-10^6$ their size, exploding into several pieces, and bending beyond 90$^{\circ}$ angles \cite{medishetty2015photosalient, naumov2013dynamic, uchida2015light}. Despite significant progress on characterizing the varied deformations in these materials {\cite{charac1-huang2021recent,charac2-awad2023mechanical}}, we do not conclusively understand the mechanisms driving the rich variety of photomechanical responses in molecular crystals.

\vspace{2mm}
\noindent Several factors are theorized to affect the macroscopic deformations of molecular crystals, including the intensity of incident light, particle geometry, and the volume changes accompanying phase transformations. For example, Naumov et al. \cite{naumov2013dynamic} synthesized over two hundred prismatic specimens of the molecular crystal, [Co(NH$_3$)$_5$(NO$_2$)]Cl(NO$_3$), and illuminated them under UV light with varying power densities. The mechanical responses of the exposed specimens ranged from rolling (at intermediate illumination) to vigorous explosions (at strong illumination). Similarly, Taniguchi et al. \cite{taniguchi2019photo} demonstrated that particle geometry has a significant impact on the mechanical response of {salicylideneamine} crystals. Under similar illumination conditions, slender, tape-like crystals of salicylideneamine bend and twist, while thicker, platelet-like crystals shear during phase transformation. In other works, researchers have also shown that significant volume changes during phase transformations (e.g., $\sim18\%$ in [Ag(2F-4spy)$_2$]BF$_4$ molecular crystal) contributes to the irreversible jumping and splitting type of deformations \cite{medishetty2015photosalient}. While these factors explain how light intensity and volume changes contributes to the elastic energy stored in the material---energy that is subsequently minimized through macroscopic deformation---they do not fully account for why some molecular crystals deform by rolling \cite{naumov2013dynamic}, while others bend or twist \cite{taniguchi2019photo} under similar light intensities.

\vspace{2mm}
\noindent Researchers have highlighted the potential role of microstructural patterns in shaping the macroscopic deformation of photomechanical crystals \cite{naumov2013dynamic, koshima2011photomechanical}. Bending, for example, is commonly observed in materials with a distinct two-phase microstructural pattern. This two-phase microstructure is shown to simultaneously stretch and contract the material, leading to bending at the macroscopic scale \cite{koshima2011photomechanical}. In contrast, a homogeneous or a solid-solution type of microstructure is often associated with explosive modes of deformation \cite{naumov2013dynamic}. Additionally, photoexcited molecules formed during isomerization reactions in molecular crystals can alter the pathways of continuum deformation \cite{photostate1:vogt2005optimal,photostate2:sanchez2000ultrafast,photostate3:yartsev1995overdamped,photostate4:waldeck1991photoisomerization, taniguchi2019photo,photo-triggered2:2024,backisomerization_time}. Though transient, these molecules can persist for extended periods before the system reverts to its equilibrium state. Besides microstructural patterns, material geometries (e.g., rod-like or film-like) and defects (e.g., cracks or pores) affect stress distributions which, in turn, can generate varied macroscopic modes of deformations \cite{naumov2020rise, naumov2013dynamic}. However, the delicate interplay between the unit cell deformation and the microstructural evolution pathways, which collectively govern the overall mechanical response, is not well understood. 

\vspace{2mm}
\noindent 
To better illustrate these mechanisms, we now focus on the salicylideneamine crystal as a representative example \cite{taniguchi2019photo}. This molecular crystal is a well-studied system in which phase transformations and the associated deformations have been extensively documented \cite{taniguchi2019photo, photo-triggered2:2024,  taniguchi2022superelasticity, wang2022domain-Salicylideneamine}. In its reference configuration, the salicylideneamine crystal consists of molecules in their enol form, periodically arranged to form a triclinic crystal system, see Fig.~\ref{fig:crystalline-order}(a). When illuminated with UV light (365~nm, 60~mWcm$^{-2}$ at room temperature), the molecules undergo an enol-to-keto photoisomerization reaction.\footnote{Salicylideneamine molecule ((S)-N-3,5-di-tert-butylsalicylidene-1-(1-naphthyl)ethylamine) is a photochromic chiral molecule that undergoes a enol-to-keto photoisomerization reaction. This reaction is accompanied by a conformational change in the molecular structure. In particular, the dihedral angle between the {salicyl and naphthyl} planes of the salicylideneamine molecule changes from $48.3^\circ$ to $47.4^\circ$, see Fig.~\ref{fig:isomerization}(c)\cite{taniguchi2019photo}.} This chemical reaction generates a surficial layer of keto-molecules and triggers the $\beta \to \gamma$ solid-to-solid phase transformation of the molecular crystal, see Fig.~\ref{fig:crystalline-order}(a-b). This $\beta \to \gamma$ transformation in salicylideneamine crystals is analogous to the displacive phase transformations observed in shape-memory alloys. In the salicylideneamine crystal, this phase transformation causes a change in the lattice symmetry from triclinic to monoclinic, without a change in the nearest-neighbor molecules.

\vspace{2mm}
\noindent The structural transformations of the unit cells, together with the nucleation of the photoexcited keto-molecules, affects the macroscopic deformation of the salicylideneamine crystal. A slender, thin tape-like geometry of this molecular crystal---when exposed to UV light---undergoes an abrupt bending and twisting deformation, see Fig.~\ref{fig:phase-transformation-slender}. A closer look at this photo-triggered $\beta \to \gamma$ transformation shows a fast and step-wise bending of the molecular crystal followed by a twist. A thicker specimen of this salicylideneamine crystal, when exposed to the exact UV light stimulus, shows a two-phase microstructure and undergoes a shearing deformation, see Fig.~\ref{fig:phase-transformation-thick} \cite{taniguchi2019photo}. In both cases, the $\beta \leftrightarrow \gamma$ transformation is reversible, and the molecular crystal gradually returns to its original form on turning off the light stimulus.

\vspace{2mm}
\noindent The multi-stage bending and twisting deformation observed in the salicylideneamine crystal is one representative case of a broader range of photomechanical deformations, in which the underlying microstructural mechanisms driving the macroscopic response remains unclear {\cite{naumovReview}}. In salicylideneamine crystal, experimentalists have proposed that the bending and twisting are caused by a combination of the metastable photo-induced keto-state and the structural transformation of crystal lattices accompanying the $\beta$ (triclinic) $\to \gamma$ (monoclinic) phase change. The lattice mismatch between the keto-layer and the molecular crystal accumulates significant elastic energy that is theorized to drive the photomechanical deformation. However, the intricate interplay between elastic and thermodynamic energy minimization, photoexcited states, and particle geometry in shaping microstructural evolution pathways and the collective macroscopic deformation is still poorly understood. 

\vspace{2mm}
\noindent To predict these complex deformations, mathematical models based on theories such as the Timoshenko bimetallic strip and Euler-Bernoulli beam bending have been proposed \cite{bendingExpParNaumov}, \cite{bendingCrystalmodelNaumov}, \cite{bendingBardeen}, \cite{bilayer1}, \cite{bilayer2}. These models typically treat the molecular crystal as a layered bimorph structure with distinct reactant and product domains, and empirically correlate the bending deformation with the distribution of phases and particle geometry  \cite{bendingBardeen, taniguchi2022superelasticity}. Other studies have explored microstructural mechanisms driving the photo-induced twisting motion in molecular crystals \cite{empTwistBardeen, zhu2011reversible}, and highlight the role of internal strain energy generated by a two-phase region in driving the twist. These mathematical descriptions provide useful insights into specific deformations modes (e.g., bending), however, they often lack the detailed information on the structural transformation of unit cells and the role of photo-induced metastable states on transformation pathways, to phenomenologically predict the complex deformations that manifest in molecular crystals.

\vspace{2mm}
\noindent In recent works, researchers \cite{maghsoodiLCEs} are developing continuum theories to predict the dynamical behavior of photoresponsive elastomers--another class of light-interactive materials. Although these elastomers differ fundamentally from molecular crystals in the arrangement of their constitutive elements,\footnote{Elastomers typically consist of disordered and cross-linked polymer chains, that differs from molecular crystals which have a highly ordered arrangement of molecules on a crystalline lattice.} the continuum models offer valuable insights into the interplay between light and deformation. For example, factors such as optical penetration depth and prestress in liquid crystal elastomers significantly affect their photomechanical behavior \cite{maghsoodi2023optical, maghsoodiLCEs}. Similarly, photoabsorption \cite{corbett2015deep} and light polarization \cite{corbett2006nonlinear} are critical in shaping the dynamic and nonlinear responses of disordered elastomers. Additionally, particle geometry and geometric instabilities have been shown to impact photomechanical responses \cite{mihai2021instabilities}. These models provide important insights into the effects of crystal geometry and illumination characteristics at the macroscopic length scale; however, they do not capture the fine-scale microstructural features that collectively determine macroscopic deformations. Recent works on elastomers have started to bridge this gap by incorporating the evolution of domain patterns, providing insights into the underlying mechanisms of actuation and shape morphing in these materials \cite{lee2023macroscopic, baiCrystal, baiNemElas}. In molecular crystals, understanding these microstructural features and lattice transformation pathways is important for accurately analyzing and predicting their macroscopic deformations.

\vspace{2mm}
\noindent In this work, we develop a continuum model---rooted in the principles of the Cauchy-Born rule, the finite deformation theory, and photoreaction kinetics---to predict the photomechanical deformation pathways in molecular crystals. A key feature of our continuum model is that, by using the lattice geometries of the two crystalline phases as inputs, we can quantitatively predict various deformations in the molecular crystal---such as bending, twisting, and shearing---as a function of particle geometry, light intensity, and reaction kinetics. These predictions align with experimental observations for salicylideneamine crystals \cite{taniguchi2019photo} and provide a computational framework for engineering molecular crystals with predictable and reversible photomechanical deformations. Furthermore, we introduce a photoexcited state, represented by a local energy well that emerges only under certain light stimulus and lowers the energy barrier for phase transformation. By minimizing the total energy of the system (including elastic, thermodynamic, and gradient contributions) on a multi-well energy landscape, we establish the nuanced interplay between photoexcited states and symmetry-breaking phase transformations in molecular crystals. 

\newpage
\section{Theory}

We develop a theoretical framework for molecular crystals that undergo reversible phase transformations under a light stimulus. Using the salicylideneamine crystal as a representative example, we derive a general continuum model that can be adapted to other molecular crystals undergoing solid-to-solid phase transformations. In this model, we construct a potential energy landscape parameterized by two scalar order parameters: a structural order parameter, $\eta$, which distinguishes the $\beta$ and $\gamma$ phases, and a photoisomerization order parameter, $c$, which distinguishes the enol and keto molecules in the crystal. The $\beta \leftrightarrow \gamma$ phase change involves a symmetry-breaking lattice transformation that we account for using a finite deformation gradient, $\mathbf{F}$. The total free energy landscape has global minima corresponding to the $\beta$ phase ($\eta = 0$) and the $\gamma$ phase ($\eta = 1$), and a local minimum representing the photoexcited keto-molecules ($c=1$) that form under specific light stimulus. By minimizing the total energy of the system across a multi-well energy landscape, driven by photoreaction kinetics, we phenomenologically predict the macroscopic deformation of molecular crystals.

\vspace{2mm}
\noindent In this work, we denote scalars using lower-case letters $a$, vectors using lower-case bold letters $\mathbf{a}$, and Rank-2 tensors using upper-case bold letters $\mathbf{A}$. 

\subsection{Multi-Lattice}\label{sec:multi-lattice}
Consider a salicylideneamine molecular crystal (hereafter referred to as the molecular crystal) occupying a region $\Omega$ in a three-dimensional Euclidean space $\mathbb{R}^3$ in the reference configuration. Let $\mathbf{x}$ be the position of a point on $\Omega$, and at each point $\mathbf{x} \in \Omega$, there are crystal lattices that define the 3D arrangement of molecules.  

\begin{figure}
    \centering
    \includegraphics[width=0.6\textwidth]{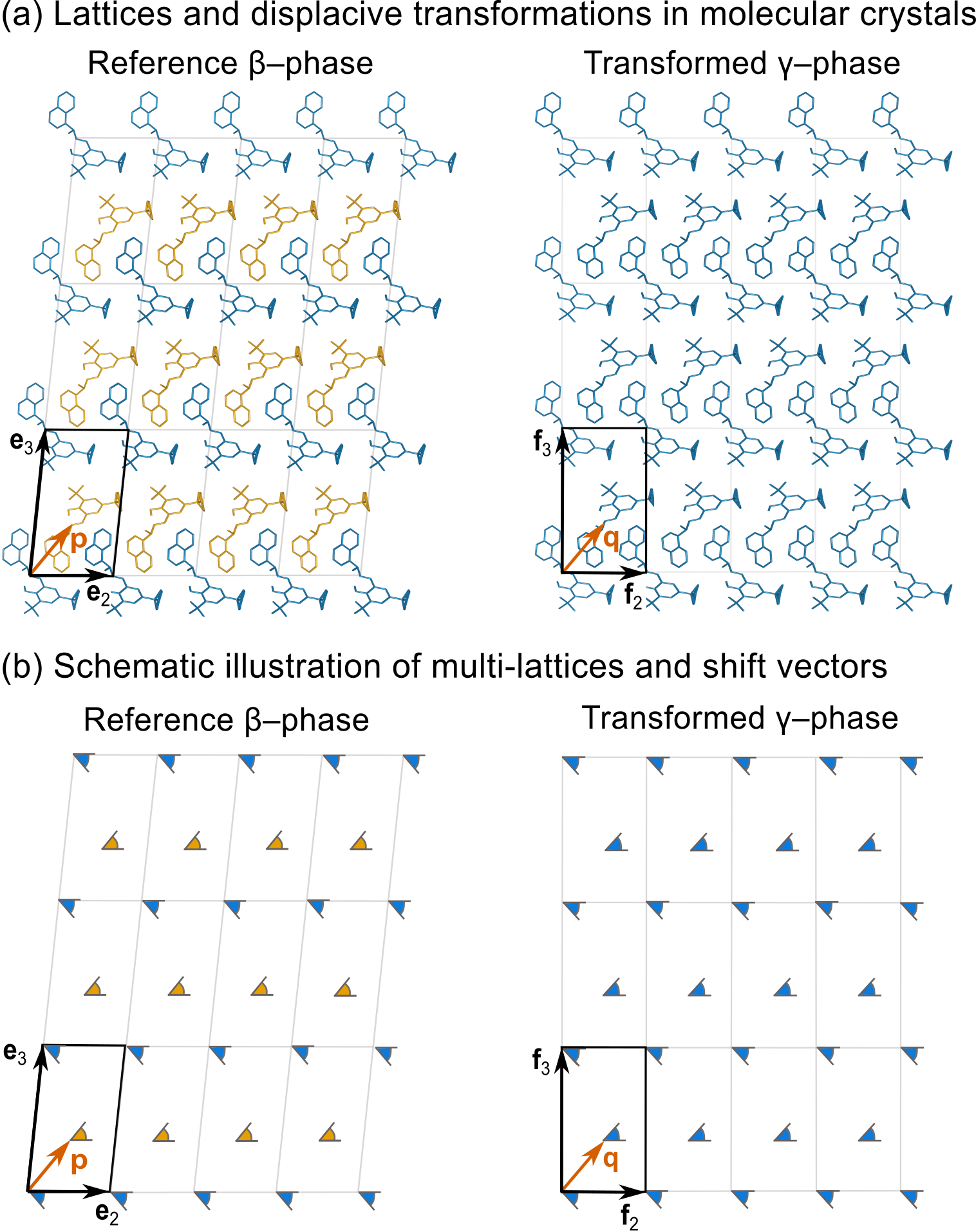}
    \caption{(a) In a salicylideneamine crystal, the molecules are periodically arranged at specific lattice sites. The blue and yellow colors respectively denote the two conformational forms of the enol-molecule, \textcolor{black}{see Fig.~\ref{fig:isomerization}(b)}. On exposure to UV light, the molecular crystal undergoes a displacive-type of phase transformation. This transformation is symmetry-breaking in which the triclinic unit cells of the reference $\beta-$phase transform into the monoclinic unit cells of the transformed $\gamma-$phase. (b) A geometric illustration of the multi-lattices in the reference and transformed phase of the salicylideneamine crystal. We denote the constituent Bravais lattice vectors in the reference and transformed phases using $\mathbf{e}_i$ and $\mathbf{f}_i$, respectively. The shift vectors in the reference and transformed phases are denoted using $\mathbf{p}$ and $\mathbf{q}$, respectively. The yellow and blue symbols highlight the change in dihedral angles during the photoisomerization reaction, see Fig.~\ref{fig:isomerization}(b).}
    \label{fig:crystalline-order}
\end{figure}

\vspace{2mm}
\noindent The lattices in {Fig.~\ref{fig:crystalline-order}} shows the molecular crystal in the reference phase (called the $\beta-$phase) containing a periodic arrangement of the enol molecules. This lattice is not a Bravais lattice---that is, we cannot describe the crystal by translating a single molecule through linearly independent lattice vectors---but is instead a collection of two congruent Bravais lattices. We refer to this arrangement as a multi-lattice and mathematically describe it using three linearly independent vectors $\{ \mathbf{e}_1, \mathbf{e}_2, \mathbf{e}_3 \}$ and a shift vector $\mathbf{p}$:

\begin{align}
 \mathcal{L}(\mathbf{e}_i, \mathbf{p}, \mathbf{o}) &= \mathcal{L}(\mathbf{e}_i, \mathbf{o}) \cup \mathcal{L}(\mathbf{e}_i, \mathbf{o}+\mathbf{p})\nonumber\\
 &= \{\mathbf{x}:\mathbf{x} = \nu^i\mathbf{e}_i + \delta \mathbf{p} + \mathbf{o} \text{ where $\nu^1, \nu^2, \nu^3$ are integers and $\delta = 0 $ or 1} \}.
 \label{MultiLattice-Reference}
\end{align}

\vspace{2mm}
\noindent In Eq.~\ref{MultiLattice-Reference}, the Bravais lattice $\mathcal{L}(\mathbf{e}_i,\mathbf{o})$ is an infinite set of points generated by translating a point $\mathbf{o}$ through three linearly independent vectors $\mathbf{e}_i$ that generates a triclinic symmetry of the $\beta-$phase. The shift vector $\mathbf{p}$ describes the offset between the two congruent Bravais lattices. 

\vspace{2mm}
\noindent Under UV light ({$\lambda = 365\mathrm{nm}$}), the enol molecules on the surface of the molecular crystal are converted to keto molecules, via a photoisomerization reaction. This reaction generates a thin layer of the photoexcited keto molecules on the crystal, which in turn triggers the abrupt $\beta \to \gamma$ phase transformation. This transformation is accompanied by the structural rearrangement of molecules in the crystal that is characteristic of a first-order displacive-type of phase transformation. The transformed phase (called the $\gamma-$phase) consists of only one type of the enol molecules arranged on a multi-lattice as shown in Fig.~\ref{fig:crystalline-order}:
\begin{align}
 \mathcal{L}(\mathbf{f}_i, \mathbf{q}, \mathbf{o}) &= \mathcal{L}(\mathbf{f}_i, \mathbf{o}) \cup \mathcal{L}(\mathbf{f}_i, \mathbf{o}+\mathbf{q})\nonumber\\
 &= \{\mathbf{x}:\mathbf{x} = \upsilon^i\mathbf{f}_i + \varepsilon \mathbf{q} + \mathbf{o} \text{ where $\upsilon^1, \upsilon^2, \upsilon^3$ are integers and $\varepsilon = 0 $ or 1} \}.
 \label{MultiLattice-Transformed}
\end{align}

\vspace{2mm}
\noindent The transformed phase can be described by three linearly independent vectors $\{\mathbf{f}_1, \mathbf{f}_2, \mathbf{f}_3\}$ generating a Bravais lattice with a monoclinic symmetry, and a shift vector $\mathbf{q}$ describes the offset between the two congruent Bravais lattices as shown in {Fig.~\ref{fig:crystalline-order}}. 

\vspace{2mm}
\noindent The multi-lattices of the $\beta$ and $\gamma$ phases respectively belong to the triclinic and monoclinic crystal families. A consequence of this material symmetry is that there are some deformations, such as rotations $\mathbf{R}$, that map a lattice back to itself. We refer to this set of rotations as a point group $\mathcal{P}$ of a lattice. For a given multi-lattice $\mathcal{L}(\mathbf{e}_i,\mathbf{p},\mathbf{o})$ of the molecular crystal we define the point group $\mathcal{P}(\mathbf{e}_i,\mathbf{p})$ as:  

\begin{align}
    \mathcal{P}(\mathbf{e}_i, \mathbf{p}) = \Bigl\{ \mathbf{R} \in SO(3) : \begin{aligned}
        &\mathbf{R}\mathbf{e}_i = \mathbf{\mu}^j_i \mathbf{e}_j \\ 
        &\mathbf{R}\mathbf{p} = \nu^i \mathbf{e}_i + \delta \mathbf{p}
    \end{aligned} 
    \text{where,} \quad \parbox{.47\linewidth}{%
    $\mu^j_i$ is a matrix of integers such that $\det[\mu^j_i] = 1$, \\ $\nu^i$ are integers, and $\delta = \pm1$ (or $+1$) if the two Bravais lattices respectively contain like (or unlike) atoms.}
    \Bigr\}
    \label{Eq:pointgroup}
\end{align}

\vspace{2mm}
\noindent From Eq.~\ref{Eq:pointgroup}, the set of rotations mapping the triclinic multi-lattice of the salicylideneamine crystal (reference $\beta-$phase) to itself is $\mathbf{R = I}$; similarly and the set of rotations mapping the monoclinic multi-lattice of the salicylideneamine crystal (transformed $\gamma-$phase) to itself is $\mathbf{R = I}$.\footnote{The set of rotations in the point group of a monoclinic multi-lattice, defined by Eq.~\ref{Eq:pointgroup}, are fewer than the set of rotations in the point group of the monoclinic Bravais lattice \cite{bhattacharya2003microstructure}. This is a consequence of the multi-lattice system and is discussed in detail in Appendix~\ref{Appendix:Multi-lattice}.} We note that these sets of rotations in the point groups of both the triclinic $\mathcal{P}(\mathbf{e}_i, \mathbf{p})$ and monoclinic multi-lattices  $\mathcal{P}(\mathbf{f}_i, \mathbf{q})$ are identical. As a result, although the $\beta \to \gamma$ phase change is a symmetry-breaking transformation, it produces only a single lattice variant during the phase transformation.

\subsection{Deformation Gradient}

\noindent We describe the deformation of the molecular crystal as a function $\mathbf{y}: \Omega \to \mathbb{R}^3$, in which $\mathbf{y(x)}$ denotes the position of the point $\mathbf{x}$ in the deformed configuration $\mathbf{y}(\Omega)$. Given any deformation $\mathbf{y(x)}$, we describe the deformation gradient $\mathbf{F(x)} = \nabla\mathbf{y(x)}$ (a Rank-2 tensor) as a matrix of partial derivatives with components,

\begin{equation}
    \mathbf{F}(\mathbf{x}) = \nabla\mathbf{y}(\mathbf{x})_{ij} = \frac{\partial y_i}{\partial x_j} \quad i,j = 1,2,3.
\end{equation}

\vspace{2mm}
\noindent In {Fig.~\ref{fig:CaucyBorn-Interface}(a)}, on illuminating the molecular crystal $\Omega$ with UV light, the constituent Bravais lattices at the position $\mathbf{x}$ in the reference phase undergo a solid-to-solid phase transformation and deform as follows:

\begin{equation}
    \mathbf{f}_i = \mathbf{A}\mathbf{e}_i.
    \label{DeformationGradient}
\end{equation}

\begin{table}[ht]
    \small
    \centering
    \renewcommand\arraystretch{1.5}
    \addtolength{\leftskip} {-2cm}
    \addtolength{\rightskip}{-2cm}
    \begin{tabular}{lll}
    \hline
     & Reference $\beta-$phase & Transformed $\gamma-$phase \\
    \hline
    Lattice parameters & $a = 6.174 $\AA & $a = 6.189 $\AA \\
     & $b = 9.912$\AA & $b = 9.871$\AA \\
     & $c = 19.602$\AA & $c = 19.741$\AA \\
     & $\alpha = 84.484^{\circ}$ & $\alpha = 90^{\circ}$ \\
     & $\beta = 86.208^{\circ}$ & $\beta = 96.946^{\circ}$ \\
     & $\gamma = 88.252^{\circ}$ & $\gamma = 90^{\circ}$ \\
    
    Symmetry &Triclinic & Monoclinic \\
    {Shift vector} & $\mathbf{p} = [-2.3, {5.0}, 6.8]$ & $\mathbf{q} = [-3.7, {4.9}, 6.7]$\\
    \hline
    \end{tabular}
    \caption{Lattice geometries of the salicylideneamine crystal in the reference $\beta-$phase and the transformed $\gamma-$phase as measured in the experiments by Taniguchi et al. ~\cite{taniguchi2019photo}. The shift vectors $\mathbf{p}$ and $\mathbf{q}$ for the reference and transformed phases are calculated using the respective CIF files \cite{taniguchi2019photo}.}
    \label{Table:lattice_parameters}
\end{table}

\noindent In Eq.~\ref{DeformationGradient}, the deformation gradient $\mathbf{A}$ uniquely maps the triclinic lattice in the reference phase $\mathcal{L}(\mathbf{e}_i, \mathbf{p}, \mathbf{o})$ to the monoclinic lattice in the transformed phase $\mathcal{L}(\mathbf{f}_i, \mathbf{q}, \mathbf{o})$. Using the structural measurements of the salicylideneamine crystal listed in Table~\ref{Table:lattice_parameters} (from Ref.~\cite{taniguchi2019photo}) and our structural transformation algorithms from {Ref.~\cite{zhang2023designing}}, we compute the deformation gradient to be:

\begin{align}
    \mathbf{A} = \begin{bmatrix}
1.002 & -0.030 & -0.186 \\
0 & 0.996 & -0.094 \\
0 & 0 & 1.006 
\end{bmatrix}
\label{FSalicyli}
\end{align}

The deformation gradient in Eq.~\ref{FSalicyli} is related to the continuum deformation of the molecular crystal via the Cauchy-Born rule \cite{Cauchy-Born-ericksen}.

\subsection{Cauchy-Born Rule}\label{sec:Cauchy-Born}
\noindent In the continuum theory of crystalline solids, the Cauchy-Born rule states that if a point $\mathbf{x}$ on a solid $\Omega$ deforms by $\mathbf{y}(\mathbf{x})$ in response to an external stimulus, the lattices at this point follow this overall deformation. Thus the deformation gradient $\mathbf{A}$ in Eq.~\ref{DeformationGradient}, representing the lattice transformation, also describes the overall deformation of the molecular crystal. The Cauchy-Born rule has been shown to accurately describe microstructural features (e.g., orientation of phase boundaries, twin interfaces) in shape-memory alloys, ferroelectrics, ferromagnets, and in this section we show that this condition can be applied to predict the orientations of coherent interfaces in molecular crystals, see Table~\ref{Table:Interface-Compatability}. 

\begin{figure}
    \centering
    \includegraphics[width=0.6\textwidth]{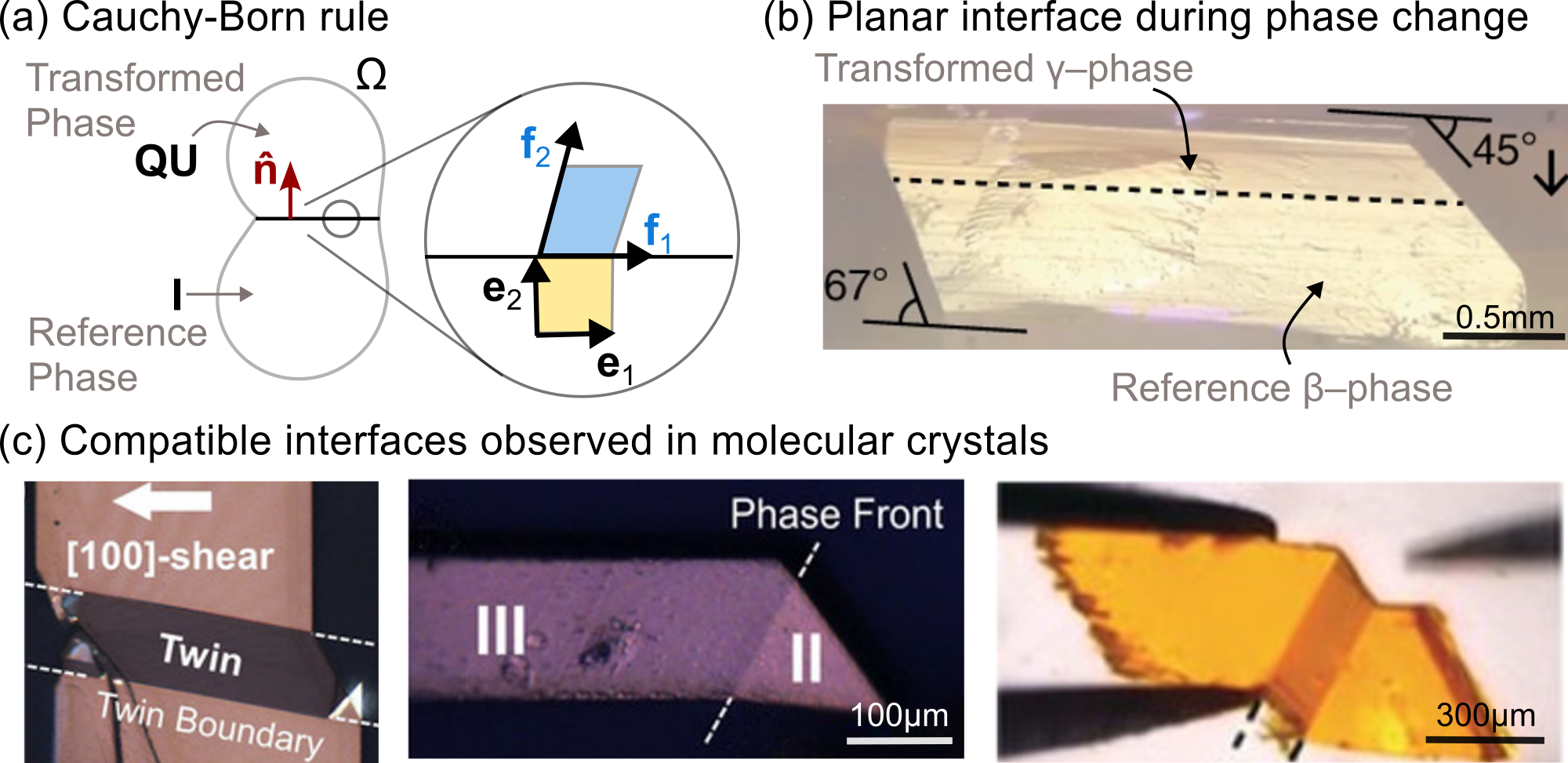}
    \caption{
    (a) A schematic illustration of a partially transformed molecular crystal $\Omega$ satisfying the Cauchy-Born rule \cite{Cauchy-Born-ericksen}. The deformation gradient of the reference phase is denoted by $\mathbf{I}$, and that of the transformed phase is denoted by $\mathbf{A=QU}$. The deformation at the interface is continuous and the two phases fit compatibly, if and only if, the deformation gradients satisfy the Hadamard jump condition on the interface.     
    (b) The salicylideneamine crystal generates a two-phase microstructure and the planar phase boundary approximately satisfies the jump condition, see {Table~\ref{Table:Interface-Compatability}}. Reprinted from Ref.~\cite{taniguchi2019photo} under a \href{https://creativecommons.org/licenses/by/4.0/}{Creative Commons Attribution 4.0 International License}. (c) Coherent and nearly compatible interfaces are observed in a variety of molecular crystals and provide evidence for the use of the Cauchy-Born rule in developing our theoretical framework, see {Table~\ref{Table:Interface-Compatability}}. In subfigure (c) the images on the left and center are reprinted with permissions from Ref.~\cite{Coh-twinTIPS} (Copyright 2020 John Wiley and Sons); the image on the right is reprinted with perimission from Ref.~\cite{Coh-twinSlideImage} (Copyright 2017 John Wiley and Sons).}
    \label{fig:CaucyBorn-Interface}
\end{figure}

\vspace{2mm}
\noindent Consider the deformation of a molecular crystal $\Omega$ in {Fig.~\ref{fig:CaucyBorn-Interface}(a)}. A part of this crystal has transformed under a light stimulus and its deformation gradient is $\mathbf{A}$, and the remainder of the crystal is in the reference configuration $\mathbf{I}$. The interface separating the two regions---phase boundary---is coherent and the jump in the  deformation gradient can be approximated by the kinematic compatibility condition \cite{bhattacharya2003microstructure}:

\begin{align}
    \mathbf{A-I=a\otimes\hat{n}},
    \label{eq: compatibility condition}
\end{align}

\noindent for vectors $\mathbf{a}, \hat{\mathbf{n}}$. Using the deformation gradient $\mathbf{A}$ describing lattice transformation at the atomic scale, the solution to Eq.~\ref{eq: compatibility condition} provides the orientation of the compatible interface, $\hat{\mathbf{n}}$, at the continuum scale. It is important to note that this tensor $\mathbf{A}$ accounts only for the deformation of the constituent Bravais lattices in predicting the geometric features of continuum microstructures. The relative movement between {constituent} Bravais lattices---represented by the change in the shift vectors---is not quantified in this deformation gradient. 

\vspace{2mm}
\noindent Using the experimental characterization data of molecular crystals undergoing solid-to-solid phase transformations {\cite{taniguchi2019photo}, \cite{Coh-twinTIPS}, \cite{Coh-twinSlideImage}, \cite{compInt-PHA}}, we compute the deformation gradients of their constituent Bravais lattices. These deformation gradients are necessary to compute the orientation of coherent interfaces or phase boundaries $\hat{\mathbf{n}}$ in Eq.~\ref{eq: compatibility condition}. {Table~\ref{Table:Interface-Compatability}} lists these interface orientations $\hat{\mathbf{n}}$ from our analytical calculations, which we then compare with experimentally reported measurements $\hat{\mathbf{m}}$. We find that our predicted orientations of the interfaces compare reasonably with experiments ($\Delta \theta < 7^\circ$), and supports the use of the Cauchy-Born rule in developing our continuum model.

\begin{table}[ht]
    \centering
    \begin{tabular}{l l l l}
    \hline
         Molecular Crystal &Theory $(\hat{\mathbf{n}})$ & Experiment $(\hat{\mathbf{m}})$ &  $ \Delta\theta$ \\
     \hline
     Salicylideneamine \cite{taniguchi2019photo} & $(0.00,0.12,0.99)$ & $(0.00,0.00,1.00)$ & $5.73^{\circ}$ \\
     &&&\\
     Triisopropylsilylethynyl & $(-0.97,-0.03,0.00)$ & $(-0.91,-0.40,0.00)$ &  $3.70^{\circ}$ \\
     pentacene \cite{Coh-twinTIPS} & & &  \\
     5-chloro-2-nitroaniline \cite{Coh-twinSlideImage} & $(-0.72,0.00,-0.68)$ & $(-0.70,0.00,-0.70)$ & $1.68^{\circ}$\\   
     &&&\\
     Palladium hexafluoroacetyl  &  $(-0.84,0.07,-0.52)$ & $(-0.79,0.00,-0.60)$ &  $6.79^{\circ}$\\
     acetonate \cite{compInt-PHA} & & &  \\
     \hline
    \end{tabular}
    \caption{A list of solutions for the compatible interfaces formed during phase transformation in select molecular crystals. The orientations of the phase boundary $\hat{\mathbf{n}}$ are calculated using Eq.~\ref{eq: compatibility condition} and compared with experimental measurements $\hat{\mathbf{m}}$ from Refs.~\cite{taniguchi2019photo, Coh-twinTIPS, Coh-twinSlideImage, compInt-PHA}. The difference in interface orientations between theoretical predictions and experimental results is calculated as $\Delta\theta = \mathrm{cos}^{-1}\frac{\mathbf{\hat{m}}\cdot{\mathbf{\hat{n}}}}{|\mathbf{\hat{m}}||\mathbf{\hat{n}}|}$.}
    \label{Table:Interface-Compatability}
\end{table}

\vspace{2mm}
\noindent The microstructures shown in Fig.~\ref{fig:CaucyBorn-Interface} are observed in single crystals subjected to quasi-static mechanical or thermal loads. For example, the twin boundaries in Fig.~\ref{fig:CaucyBorn-Interface}(c) form under shear loads applied over extended periods of time ($\sim1\mathrm{min}$) and are reported to be stable at intermediate loading stages ~\cite{Coh-twinTIPS, Coh-twinSlideImage}. Under these quasi-static conditions, we assume that the shift vectors in individual unit cells adjust locally for any given deformation and minimize the overall energy of the system. Similarly, in the salicylideneamine crystal {(Fig.~\ref{fig:CaucyBorn-Interface}(b))}, the in-plane strains arising from the photo-triggered keto molecules drives the $\beta \to \gamma$ phase change. Taniguchi et al. \cite{taniguchi2019photo} imaged this photo-triggered phase change in salicylideneamine over extended time scales ($\sim 0.1-1\mathrm{s}$) and reported the formation of two-phase microstructures as a consequence of total energy minimization in the molecular crystal. For these systems, the elastic energy of a multi-lattice material can be described solely as a function of the deformation gradient, with the shift vector minimized out of the problem \cite{james1987shiftvector, bhattacharya2003microstructure, bhattacharya1997relaxation} (see Appendix~\ref{Appendix:Multi-lattice} for further details). With these assumptions in place, we proceed to construct the free energy density for the salicylideneamine crystal.

\subsection{Free Energy Landscape}\label{sec:FreeEnergyDensity}
\noindent We define the free energy of the system $\psi$---as a function of a structural order parameter $\eta$, a photoisomerization order parameter $c$, and the deformation gradient $\mathbf{F}$---with the energy wells corresponding to the $\beta$ ($\eta = 0$) and the $\gamma$ ($\eta = 1$) phases, respectively. Under UV light (of wavelength $\lambda=365\mathrm{nm}$), a local minimum corresponding to the photo-triggered keto-phase $c=1$ forms in the continuum energy landscape.

\vspace{2mm}
\noindent Following Refs.~\cite{james1987shiftvector, bhattacharya2003microstructure, bhattacharya1997relaxation}, the energy of a multi-lattice system can be described as a function of both the deformation gradient and the shift vector $\tilde{\psi}(\mathbf{F}\mathbf{e}_i,\mathbf{p},\eta, c)$. However, for equilibrium calculations, we can minimize the shift out of the problem by assuming the existence of a function $\tilde{\mathbf{p}}(\mathbf{F}\mathbf{e}_i,\eta, c)$ such that $\tilde{\psi}(\mathbf{F}\mathbf{e}_i,\mathbf{p},\eta, c) \geq \tilde{\psi}(\mathbf{F}\mathbf{e}_i,\tilde{\mathbf{p}}(\mathbf{F}\mathbf{e}_i,\eta, c),\eta,c)$ \cite{james1987shiftvector, bhattacharya2003microstructure, bhattacharya1997relaxation}. This leads to a free energy function $\psi(\mathbf{F},\eta,c) = \tilde{\psi}(\mathbf{F},\tilde{\mathbf{p}},\eta,c)$ that solely depends on the deformation gradient for a given microstructural state \cite{james1987shiftvector, bhattacharya2003microstructure, bhattacharya1997relaxation} (see the Appendix~\ref{Appendix:Multi-lattice} for further details).

\vspace{2mm}
\noindent We assume that $\psi$ is defined and continuous for all $\mathbf{F}\in\mathcal{D} = \{\mathbf{F} \in \mathrm{M}^{3 \times 3} | \mathrm{det}\mathbf{F} > 0\}$ and for all $\eta$ in the neighborhood $0 \le \eta \le 1$. Here, $\mathrm{M}^{3\times3}$ denotes a set of real $m\times n$ matrices and the order parameter $\eta = 0$ and $\eta = 1$ corresponds to the reference and transformed phases of the molecular crystal, respectively. We assume that the free energy is Galilean invariant for all $\mathbf{F}\in\mathcal{D}$, for all $\eta$ in the neighborhood $0 \le \eta \le 1$, and each orthogonal rotation tensor $\mathbf{R}$ with $\mathrm{det}\mathbf{R}=1$,

\begin{equation}\label{Eq:Frame-indifference}
    \psi(\mathbf{RF},\eta, c) = \psi(\mathbf{F},\eta, c).
\end{equation}

\noindent Additionally, the material symmetry argument requires that any two sets of lattice vectors generating the same unit cell must have the same free energy. We combine the material-symmetry and frame-indifference arguments and confine the rotations $\mathbf{R}$ to the point group $\mathcal{P}(\mathbf{e}_i, \mathbf{p})$ to obtain:

\begin{equation}
    \psi(\mathbf{RFR}^{\top},\eta, c) = \psi(\mathbf{F},\eta,c)~\forall~\mathbf{R}\in\mathcal{P}(\mathbf{e}_i, \mathbf{p}).
    \label{invariance}
\end{equation}

\noindent The invariance restriction in Eq.~\ref{invariance} implies that the free energy is a function of a symmetric tensor and we construct the energy density assuming isotropic material properties. This idealization leads to:
\begin{equation}
    \psi(\mathbf{F},\eta,c) = \psi({\mathbf{E}},\eta,c).
    \label{Green-Lagrange Strain}
\end{equation}

\noindent In Eq.~\ref{Green-Lagrange Strain}, we introduce a Green-Lagrange strain tensor $\mathbf{E}$  that is related to the deformation gradient as follows:

\begin{align}
    {\mathbf{E}} &= \frac{1}{2}({\mathbf{F}}^{\top}{\mathbf{F}}-\mathbf{I})\label{GLstrain}
\end{align}

\noindent Or equivalently, using Eqs.~\ref{invariance}-\ref{GLstrain} the free energy function satisfies

\begin{equation}
    \psi(\mathbf{R{\mathbf{E}}}\mathbf{R}^{\top},\eta,c) = \psi({\mathbf{E}},\eta,c)~\forall~\mathbf{R}\in \mathcal{P}(\mathbf{e}_i, \mathbf{p}).
    \label{GLstrain-invariance}
\end{equation}

\noindent Eq.~\ref{GLstrain-invariance} holds for all rotations in the finite point group of the undistorted crystalline lattice $\mathcal{P}\mathbf{(e}_i,\mathbf{p})$ and ${\mathbf{E}} \in \mathcal{D}^{\mathrm{S}} \subset \mathcal{D}$. We denote $\mathcal{D}^{\mathrm{S}}$ to be a subset of $\mathcal{D}$ consisting of positive definite symmetric matrices. Eq.~\ref{GLstrain-invariance} requires that the free energy $\psi$ and the strain tensor ${\mathbf{E}}$ are invariant under material symmetry transformations.

\begin{figure}
    \centering
    \includegraphics[width=1\textwidth]{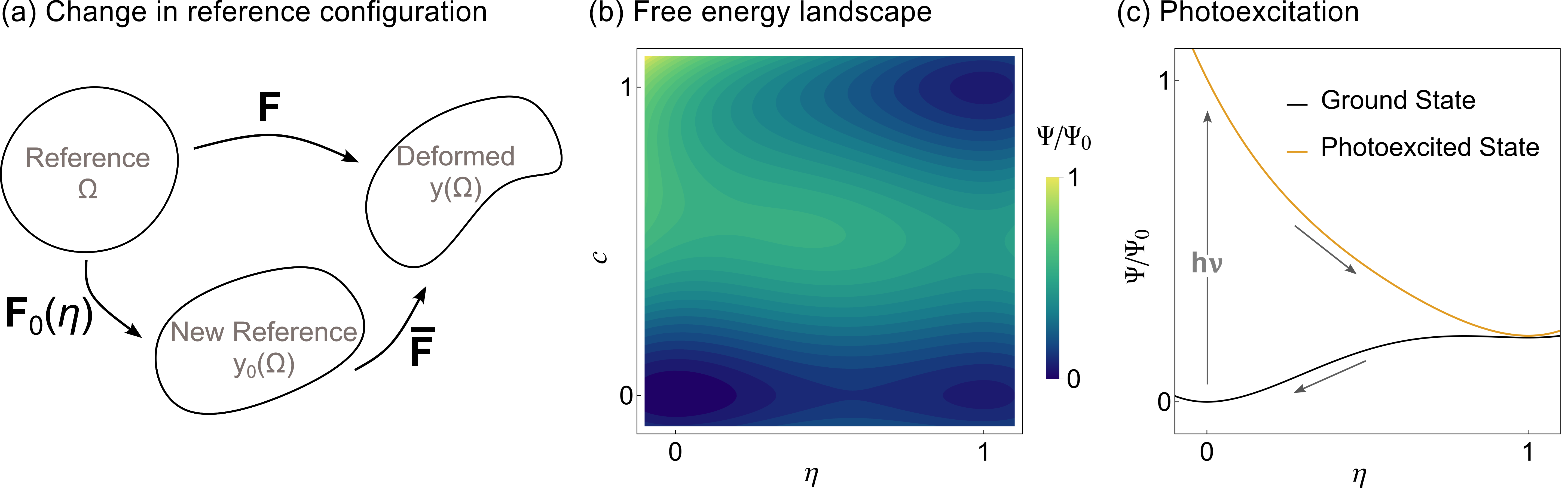}
    \caption{(a) A schematic illustration of the molecular crystal in its initial reference configuration $\Omega$, the stress-free or the new reference configuration $\mathbf{y}_0(\Omega)$, and the deformed configuration $\mathbf{y}(\Omega)$. The deformation gradients are related by $\mathbf{F}=\bar{\mathbf{F}}\mathbf{F}_0$. (b) A contour plot of the free-energy $\Psi$ as a function of the structural order parameter $\eta$ and the photoisomerization order parameter $c$. The structural order parameter $\eta$ distinguishes the triclinic $\beta-$phase from the monoclinic $\gamma-$phase, while the photoisomerization order parameter $c$ distinguishes between the enol and keto forms of the molecule. 
    (c) A simplified representation of the free energy landscape related to photoexcitation. Illuminating a salicylideneamine crystal with UV light $\lambda = 365~\textrm{nm}$ induces a barrierless enol-to-keto photoisomerization reaction. This photoexcitation reduces the energy barrier governing the solid-to-solid $\beta\to\gamma$ phase transformation in the molecular crystal. The energy barrier corresponding to the $\beta$ and $\gamma$ phases is $\approx 5.48\, \text{kJ}{\text{m}}^{-3}$ and is fitted with the calorimetric measurements reported in Ref.~\cite{taniguchi2019photo}.}
    \label{fig:MultDecomp-FreeEnergy}
\end{figure}

\vspace{2mm}
\noindent The total deformation gradient of the molecular crystal $\mathbf{F}$ relative to its reference configuration $\Omega$ can be decomposed as illustrated in Fig.~\ref{fig:MultDecomp-FreeEnergy}(a): 

\begin{align}
    \mathbf{F}=\bar{\mathbf{F}}\mathbf{F}_0
    \label{eq: Total deformation gradient}
\end{align}
First, the deformation gradient $\mathbf{F}_0(\eta)$ characterizes the spontaneous deformation of the salicylideneamine crystal during the $\beta \to \gamma$ phase change. This deformation gradient $\mathbf{F}_0(\eta)$ maps the reference configuration of the molecular crystal $\Omega$ (corresponding to the $\beta-$phase) to a new configuration $\mathbf{y}_0(\Omega)$ (corresponding to the $\gamma-$ phase); see Fig.~\ref{fig:MultDecomp-FreeEnergy}(a). The new reference configuration represents a stress-free state, and any deformation away from this configuration introduces a finite elastic energy into the system. We represent these deviations from the stress-free configurations using the deformation gradient $\bar{\mathbf{F}}$. For future use, we define the symmetric strain tensors $\mathbf{E}_0 = \frac{1}{2}(\mathbf{F}_0^\top\mathbf{F}_0 - \mathbf{I})$ and $\bar{\mathbf{E}} = \frac{1}{2}(\bar{\mathbf{F}}^\top\bar{\mathbf{F}} - \mathbf{I})$, respectively.

\vspace{2mm}
\noindent In Eq.~\ref{eq: Total deformation gradient}, we construct the generalized spontaneous deformation gradient ${\mathbf{F}}_0(\eta)$ by assuming a linear coupling between the deformation gradient and the structural order parameter $\eta$:

\begin{align}
    {\mathbf{F}}_0(\eta) &= \eta [\mathbf{A} -\mathbf{I}]+\mathbf{I} 
    \label{SpontaneousF}
\end{align}
 
\noindent Eq.~\ref{SpontaneousF} describes an affine lattice deformation between the reference$-\beta$ (triclinic) and transformed$-\gamma$ (monoclinic) phases. In the reference configuration with $\eta = 0$, the spontaneous deformation gradient is an identity tensor ${\mathbf{F}}_0(\eta=0) = \mathbf{I}$. In the deformed configuration with $\eta = 1$, the spontaneous deformation gradient maps the {triclinic-to-monoclinic distortion} of the unit cells ${\mathbf{F}}_0(\eta = 1) = \mathbf{A}$. These deformation gradients $\mathbf{F}_0(\eta = 0)$ and $\mathbf{F}_0(\eta=1)$ are used to construct the spontaneous strain tensor $\mathbf{E}_0 = \frac{1}{2}(\mathbf{F}_0^{\top}\mathbf{F}_0-\mathbf{I})$ and correspond to the energy-minimizing wells in {Fig.~\ref{fig:MultDecomp-FreeEnergy}(b)}. 

\vspace{2mm}
\noindent We postulate the total free energy function $\Psi(\mathbf{E},\eta, c)$ at constant temperature as:

\begin{equation}
    \Psi({\mathbf{E}},\eta, c) = \int_\Omega [\frac{1}{2}\nabla\eta\cdot\mathbf{K}_1\nabla\eta + \frac{1}{2}\nabla c\cdot\mathbf{K}_2\nabla c + \psi({\mathbf{E}},\eta,c)]\mathrm{d}\mathbf{x}.
    \label{Eq:free-energy}
\end{equation}
    
\vspace{2mm}
\noindent In Eq.~\ref{Eq:free-energy}, $\mathbf{K}_1,\mathbf{K}_2\in\mathrm{M}^{3\times 3}$ is a positive definite matrix and the free energy density $\psi({\mathbf{E}},\eta,c)$ satisfies both the frame indifference and material symmetry properties, see Eq.~\ref{GLstrain-invariance}. Specifically, we construct the total free energy as:

\begin{align}
    \Psi({\mathbf{E}},\eta,c) &= \int_\Omega \bigl\{ \underbrace{\frac{1}{2}\nabla c\cdot\mathbf{K}_1\nabla c + \frac{1}{2}\nabla\eta\cdot\mathbf{K}_2\nabla\eta}_{\mathrm{gradient}\,\mathrm{energy}} 
    + \underbrace{A_1 [\eta^2 (1-\eta)^2] + A_2 \left[\frac{\eta^2}{2}-\frac{\eta^3}{3}\right]}_{\beta\,  \leftrightarrow\, \gamma\, \mathrm{thermodynamic}\,\mathrm{energy}} \nonumber\\   
    & + \underbrace{A_3[c^2(1-c)^2] + A_4(1-\eta^2) c^2}_{\mathrm{enol\, \leftrightarrow\, keto\, isomerization}\,\mathrm{energy}} + \underbrace{\frac{1}{2}J_0 \bar{\mathbf{E}}:\mathbb{C}{\bar{\mathbf{E}}}}_{\mathrm{elastic}\,\mathrm{energy}}-\underbrace{\mathbf{\sigma}_0(c):\mathbf{E}}_{\mathrm{mechanical}\,\mathrm{load}}\bigr\} \, \mathrm{d}\mathbf{x}.
\label{Eq:freeEnergy}
    \end{align}

\vspace{2mm}
\noindent {Fig.~\ref{fig:MultDecomp-FreeEnergy}(b)} shows a plot of the multi-well free energy function corresponding to the salicylideneamine crystal. In Eq.~\ref{Eq:freeEnergy}, the gradient energy term penalizes changes in the order parameter $\eta$ and we assume an isotropic form of the gradient energy coefficients $\mathbf{K}_1=\kappa_1\mathbf{I}$ and $\mathbf{K}_2=\kappa_2\mathbf{I}$. The thermodynamic energy corresponds to the enthalpy of the $\beta \to \gamma$ phase transformation, in which the first term describes a double-well potential, and the second term introduces an energy barrier associated with the structural phase change. This energy barrier for $\beta \leftrightarrow \gamma$ phase transition was measured by Taniguchi et al. \cite{taniguchi2019photo} using differential scanning calorimetry and is $0.20\, \text{kJ}\, {\text{mol}}^{-1}$ $\approx 5.48\, \text{kJ}{\text{m}}^{-3}$. We fit the coefficients $A_1, A_2$ with the differential scanning calorimetry (DSC) measurements of the molecular crystal and the thermodynamic energy polynomial describes the multi-well landscape with an energy barrier of $\approx 5.48\, \text{kJ}{\text{m}}^{-3}$ at room temperature, see Fig.~\ref{fig:MultDecomp-FreeEnergy}(b,) \cite{taniguchi2019photo, taniguchi2022superelasticity}. The coefficients $A_3, A_4$ correspond to the energy barriers governing the enol-to-keto isomerization of the salicylideneamine crystal. At room temperatures, the enol phase ($c=0$) is energetically favorable, however, under UV light ($\lambda = \mathrm{365nm},~\mathcal{I} = \mathrm{60 mWcm^{-2}}$ \cite{taniguchi2019photo}) the photoreaction kinetics drives the formation of keto molecules $(c=1)$.

\vspace{2mm}
\noindent The elastic energy penalizes deformations not conforming to the stress-free states (i.e., corresponding to the bottom of the energy wells in {Fig.~\ref{fig:MultDecomp-FreeEnergy}(b)}). This elastic energy is a function of the Green-Lagrange strain ${\bar{\mathbf{E}}}$, the stiffness tensor $\mathbb{C}$, and $J_0 = \mathrm{det}\mathbf{F}_0$.\footnote{$J_0 = \mathrm{det}\mathbf{F}_0$ is a ratio of the volume of the molecular crystal in the new reference state with respect to its initial state, and is introduced in the elastic energy expression to account for the change in reference configuration.} Note that the strain tensors $\bar{\mathbf{E}}, \mathbf{E}$ are, respectively, the Green–Lagrange strain measures of the deformation gradients $\bar{\mathbf{F}}, \mathbf{F}$, and are related as given in Eq.~\ref{eq: Total deformation gradient}.
In our calculations, we assume isotropic elasticity and that the elastic constants of the molecular crystal remain unchanged during phase transformation. We use the elastic constants measured in experiments by Taniguchi et al. \cite{taniguchi2019photo, taniguchi2022superelasticity} to define the isotropic elasticity tensor as follows:

\begin{equation}
    \mathbb{C}=\mathrm{G}\left(\delta_{i k} \delta_{j l}+\delta_{i l} \delta_{j k}\right)+\frac{2 \mathrm{G} \nu}{1-2 \nu} \delta_{i j} \delta_{k l}.
    \label{Eq:elasticityTensor}
\end{equation}

\noindent In Eq.~\ref{Eq:elasticityTensor}, $\delta_{ij}$ is the Kronecker delta, $\mathrm{G}$ is the Shear modulus, and $\nu$ is the Poisson's ratio. Using this isotropic form of the elasticity tensor, we rewrite the elastic energy contribution as:

\begin{align}
 \frac{1}{2}J_0\bar{\mathbf{E}}:\mathbb{C}\bar{\mathbf{E}} 
& = GJ_0\left[\bar{\mathbf{E}}: \bar{\mathbf{E}}+\frac{v}{1-2 v}\left(\operatorname{tr} \bar{\mathbf{E}}\right)^2\right]\label{EQ:CP energy}
\end{align}
\noindent Finally, in Eq.~\ref{Eq:freeEnergy}, the stresses $\sigma_0(c)$ correspond to the misfit strains caused by the photoisomerized keto-molecules produced at the irradiated surface of the crystal. Taniguchi et al. \cite{taniguchi2019photo}, report a thin layer of keto-molecules to form on the molecular crystal surface ($\sim5\%$ of the crystal depth $d$). These molecules are stabilized only under UV light and generate in-plane compressive stresses of $ \approx {1.5} \mathrm{MPa}$ in the crystal (see Fig.~\ref{fig:stress Thin Plate}(b)). We model this effect as an applied mechanical load, $\sigma_0(c) = c~\mathbb{C}~[\delta\hat{\mathbf{e}_2}\otimes\hat{\mathbf{e}}_2]$, with $\delta = -0.025$ representing the in-plane misfit strains induced by the keto layer and $c$ representing the photoisomerization order parameter.

\vspace{2mm}
\noindent We numerically solve this continuum model in a finite element framework, and the specific values of the material parameters are listed in Table~\ref{Table:model-parameters} of the Appendix~\ref{Appendix:EnergyNorm-ParameterTable}.

\subsection{Kinetics and Equilibrium}\label{sec:Kinetics}
We next solve the kinetics of phase transformation in the molecular crystal at a constant temperature. We assume two key processes, namely, a photoisomerization reaction that converts the enol-to-keto molecules under UV illumination, and a domino effect-like transformation kinetics that accompanies the $\beta \to \gamma$ phase change.\footnote{Taniguchi et al. \cite{taniguchi2019photo} report a domino-effect type phase transformation in salicylideneamine crystals, caused by the conformational change of individual molecules in the crystal.} In the absence of UV illumination, the energy minimization process returns the $\gamma$ phase to the initial $\beta$ phase. We assume that the kinetics of phase transformation is much slower than the speed of sound through the molecular crystal, and thus the material is always in mechanical equilibrium.

 \vspace{2mm}
\noindent We first compute the evolution of the photoisomerization order parameter $c$ as:

\begin{align}
\textcolor{black}{\frac{\partial c}{\partial t}} & \textcolor{black}{=-\mathcal{M}_1\frac{\delta \Psi}{\delta c} + \Upsilon \mathcal{I}f }   \label{Eq:isomerizationKinetics}  
\end{align}

\vspace{2mm}
\noindent In Eq.~\ref{Eq:isomerizationKinetics}, the first term corresponds to the minimization process on the continuum energy landscape, and the second term corresponds to the linear photoisomerization reaction kinetics \cite{KineticsClassic, reactionKineticsBardeen, chemicalkineticsbook}. In Eq.~\ref{Eq:isomerizationKinetics}, $\Upsilon$ is a positive constant calibrated based on absorbance and quantum yield for a given wavelength of incident light, $\mathcal{I}$ is the light intensity, and $f$ is volume fraction of the reactant molecules in the mixture, i.e., for enol-to-keto isomerization reaction $f=(1-c)$. Following the Beer-Lambert law \cite{BeerLambertLaw}, we model the intensity of the incident light $\mathcal{I}$ to exponentially decay through the thickness of the molecular crystal {${d}$}:
 
\begin{equation}
 \mathcal{I} = \mathcal{I}_0 \mathrm{e}^{-\mu {d}}.
 \label{Eq:BeerLambert}
\end{equation}

\begin{figure}
    \centering
    \includegraphics[width=0.95\textwidth]{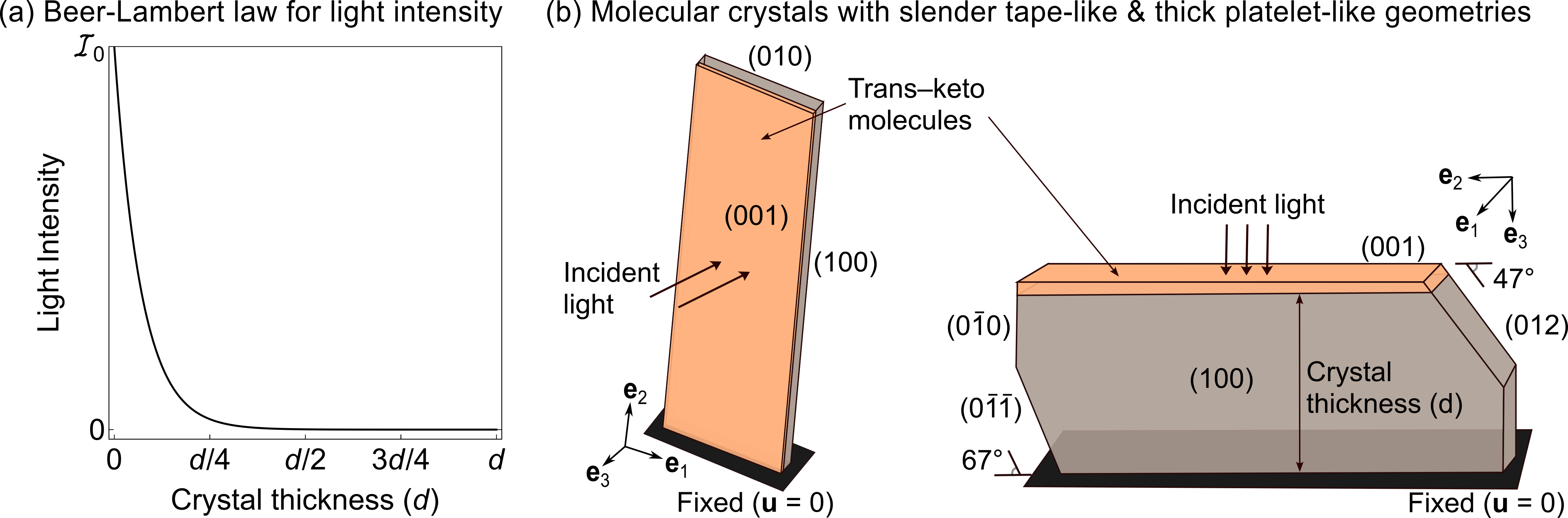}
    \caption{(a) The intensity of the applied UV light decays exponentially through the thickness of the molecular crystal. This decay in intensity $\mathcal{I}_0$ follows a Beer-Lambert law {\cite{BeerLambertLaw}}. (b) A schematic illustration of the molecular crystal geometries---slender tape-like and thick platelet-like particles---used in our calculations. The bottom face of the particles are rigidly bound to a substrate ($\mathbf{u}=0$) and the UV light is incident on the (001) plane. Under UV radiation, a thin layer of trans-keto molecules ($\sim5\%$ of crystal thickness) forms on the molecular crystal \cite{taniguchi2019photo}, inducing a compressive in-plane strain.}
    \label{fig:LightDecay-PartGeometry}
\end{figure}

\noindent Fig.~\ref{fig:LightDecay-PartGeometry}(a) shows that the intensity of the incident light is maximum at the molecular crystal surface and drives the enol-keto isomerization reaction. In Eq.~\ref{Eq:BeerLambert}, $\mathcal{I}_0$ corresponds to the intensity of light at the irradiated surface, and $\mathcal{I}$ is the attenuated intensity observed at a penetration depth $d$. The standardized absorption coefficient $\mu$ for the salicylideneamine crystal is reported by Taniguchi et al. \cite{taniguchi2019photo}.

\vspace{2mm}
\noindent We compute the $\beta \leftrightarrow \gamma$ structural transformation of the molecular crystal using:

\begin{align}
\frac{\partial \eta}{\partial t} & = -\mathcal{M}_2\frac{\delta \Psi}{\delta \eta}  + \chi (1-\eta) \nonumber \\
& = -\mathcal{M}_2 \left[ \frac{\partial \Psi}{\partial \eta} - \nabla . \left(\frac{\partial \Psi}{\partial \nabla \eta} \right)\right] + \chi (1-\eta)  \label{Eq:kinetics}
\end{align}

\vspace{2mm}
\noindent The first term in Eq.~\ref{Eq:kinetics} corresponds to the variational derivative of the free energy, and $\mathcal{M}_2$ is the mobility constant, which we calibrate using the thermal relaxation times of the salicylideneamine crystal \cite{taniguchi2019photo}.\footnote{For a more rigorous treatment of the thermal relaxation process it is important to define the mobility tensor as a function of temperature and evolve the system using thermal relaxation kinetics.} The second term represents a linearized phase transformation kinetics that models the domino cascade effect of the $\beta \to \gamma$ transformation, as reported in experiments \cite{taniguchi2019photo}. We calibrate the kinetic coefficient \(\mathcal{\chi}\) based on experimentally observed transformation kinetics and as a function of the light intensity \cite{taniguchi2019photo}. Specifically, \(\mathcal{\chi} = \mathcal{\chi}_{0} \left( 1 + 0.05 \frac{\mathcal{I}}{\mathcal{I}_{0}} \right)\), in which \(\mathcal{\chi}_{0} = 0.025\) and \(\mathcal{I}_{0} = 20 \, \text{mW} \, \text{cm}^{-2}\).

\vspace{2mm}
\noindent The molecular crystal satisfies the mechanical equilibrium condition throughout the phase transformation process, and we solve:
\begin{equation}
\mathrm{div}\left(\mathbf{P}\right)= 0.
\label{eq:mechanical-equilibrium}
\end{equation}

\noindent In Eq.~\ref{eq:mechanical-equilibrium}, $\mathbf{P}$ denotes the first Piola-Kirchhoff stress (relates forces in the deformed configuration to areas in the reference configuration) and is defined as:

\begin{equation}
    \mathbf{P} = \frac{\partial \Psi{_R}}{\partial \mathbf{F}}.
    \label{eq:pk1derivation}
\end{equation}

\noindent In Eq.~\ref{eq:pk1derivation}, $\mathbf{F}$ is the  deformation gradient. 
The total free energy $\Psi$, however, is defined as a function of the symmetric Green-Lagrange strain tensor $\bar{\mathbf{E}}$ (see Eq.~\ref{GLstrain}) and the subscript `$R$' denotes the free energy defined in the reference configuration. We, therefore, introduce additional steps in computing the first Piola–Kirchhoff stress and the mechanical equilibrium condition. 

\vspace{2mm}
\noindent First, we note that the first Piola–Kirchhoff stress is related to the Cauchy stress tensor $\bm{\sigma}$ (stresses defined in the deformed configuration) and the Jacobian determinant $J = \mathrm{det}\mathbf{F}$ as follows:

\begin{equation}
    \mathbf{P} = J \bm{\sigma} \mathbf{F}^{-\top}.
    \label{eq:pk1-cauchy}
\end{equation}

\noindent The first Piola–Kirchhoff stress in Eq.~\ref{eq:pk1-cauchy} need not be symmetric and for our numerical calculations we define a symmetric stress tensor associated with the new reference configuration:

\begin{equation}
    \mathbf{S} = \bar{J}\, \bar{\mathbf{F}}^{-1} \bm{\sigma} \bar{\mathbf{F}}^{-\top}
    \label{eq:pk2-cauchy}
\end{equation}

\noindent The symmetric stress in Eq.~\ref{eq:pk2-cauchy} is referred to as the second Piola–Kirchhoff stress and accounts for deviations from the stress-free state. From \ref{eq:pk1derivation}, \ref{eq:pk1-cauchy}, \ref{eq:pk2-cauchy}, we note that the second Piola–Kirchhoff stress can be defined as:
\begin{equation}
    \mathbf{S} = \frac{1}{{J}_0}\left(\frac{\partial \Psi_R}{\partial \bar{\mathbf{E}}}\right)\label{EQ:PK2-derivation}
\end{equation}
in which $\bar{\mathbf{E}}$ is the Green-Lagrange strain tensor and ${J}_0 = \mathrm{det}\mathbf{F}_0$. Using Eq.~\ref{Eq:freeEnergy}, Eq.~\ref{Eq:elasticityTensor},  and Eq.~\ref{EQ:PK2-derivation} we now write the second Piola–Kirchhoff stress as:
\begin{equation}
    \mathbf{S} = 2 G \, \bar{\mathbf{E}} + 2 G \frac{v}{1-2 v}\left(\operatorname{tr} \bar{\mathbf{E}}\right) \mathbf{I} - \frac{1}{{J}_0} \mathbf{F}_0 \mathbf{\sigma}_0(c) \mathbf{F}_0^{\top}.\label{EQ:PK2-formual}
\end{equation}

\noindent Finally, using Eqs.~\ref{eq:pk1-cauchy}-\ref{eq:pk2-cauchy}, we write the first Piola–Kirchhoff stress as:
\begin{equation}
    \mathbf{P} = {J}_0 \mathbf{F} \mathbf{F}_0^{-1} \mathbf{S} \mathbf{F}_0^{-\top}
\end{equation}

\noindent We numerically solve these stresses and the governing equations in a finite element framework. Specifically, we consider mechanical deformation in two types of particle geometries: a slender tape-like and a thick platelet-like molecular crystal, see {Fig~\ref{fig:LightDecay-PartGeometry}(b)}. Further details on the numerical implementation of the theory are described in detail in the {Appendix~\ref{Appendix:Weakform}}.

\section{Results}

Our results highlight the interplay between the structural transformation of unit cells, photoexcited states, and energy-minimizing pathways in governing the macroscopic deformation of the molecular crystal. We establish a continuum model---which using lattice deformation at the atomic scale as a primary input---phenomenologically predicts the diverse and complex deformation pathways of molecular crystals (e.g., bending, twisting, shearing). By incorporating the thermodynamic energy barriers and photoisomerization rates, our model also captures the phase transformation kinetics in molecular crystals.
These results quantitatively align with the experimental measurements of deformation types observed in the salicylideneamine crystal with specific particle geometries and incident light conditions \cite{taniguchi2019photo}. The developed framework is general and can be adapted to other molecular crystals undergoing a solid-to-solid phase transformation under a light stimulus.
In general, the results illustrate a potential application of our modeling framework to theoretically guide the design and actuation of molecular crystals with reversible deformation.

\subsection{Photomechanical Deformation} \label{sec:deformarion-slenderCrystal}

\begin{figure}
    \centering
    \includegraphics[width=0.65\textwidth]{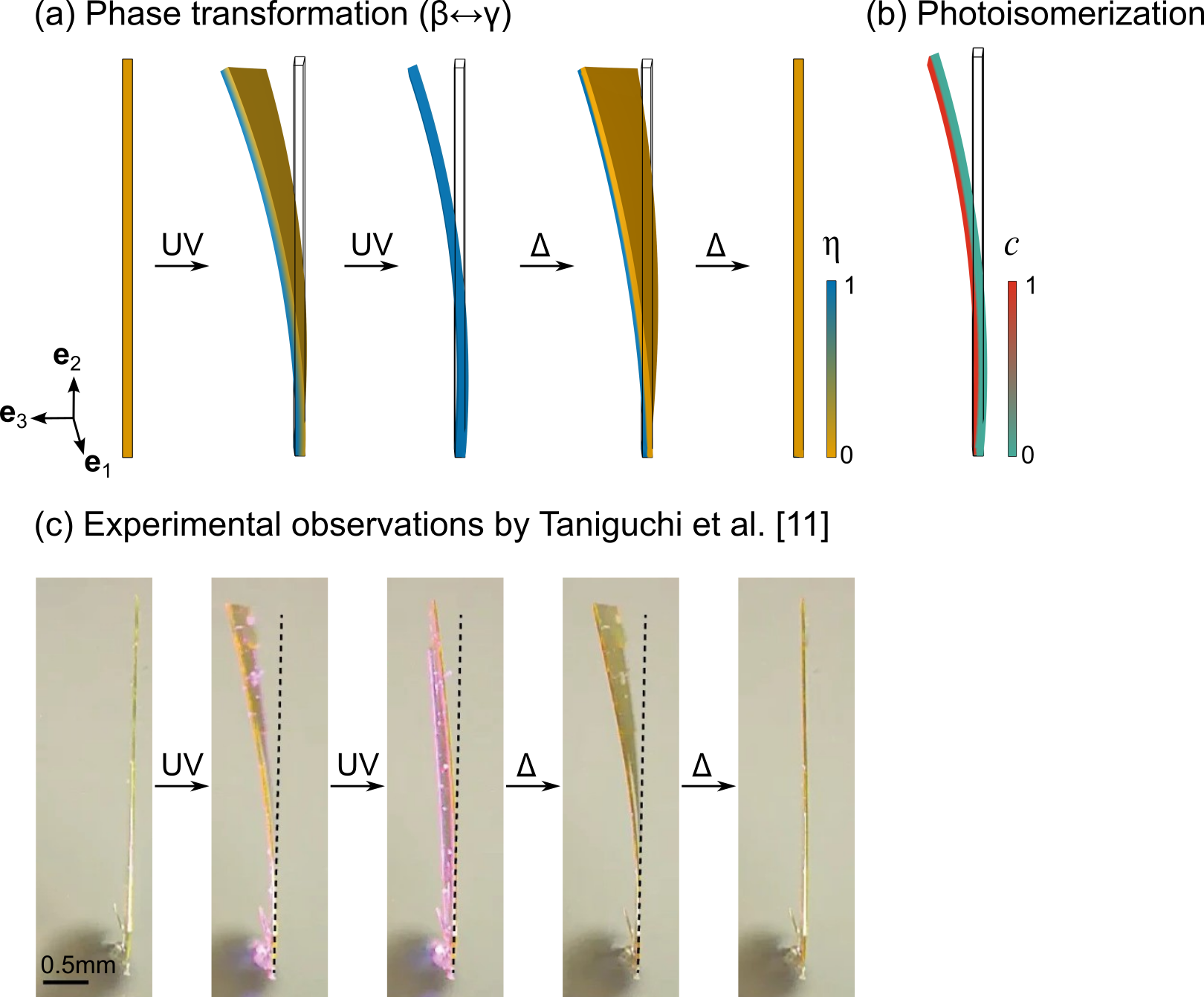}
    \caption{(a,b) Theoretical predictions of the photomechanical deformation of a slender salicylideneamine crystal. (a) On UV irradiation, the molecular crystal undergoes a $\beta \to \gamma$ phase transformation that is accompanied by a stepwise bending and twisting deformation of the crystal. On turning off the light stimulus, the molecular crystal returns to its initial configuration {(see $\Delta$)}. The color bar indicates the extent of phase transformation, with $\eta = 0$ and $\eta = 1$ corresponding to the triclinic and monoclinic phases, respectively. (b) The photoisomerization reaction generates a thin surface layer of keto-molecules that are metastable (i.e., existing only under UV radiation) and are shown in the fully transformed $\gamma-$phase of the molecular crystal. The color bar represents the concentration of keto molecules in the crystal. (c) Our theoretical predictions closely match the experimental observations of the stepwise bending and twisting deformations of the salicylideneamine crystal, as reported by Taniguchi et al. \cite{taniguchi2019photo}. Reprinted from Ref.~\cite{taniguchi2019photo} under a \href{https://creativecommons.org/licenses/by/4.0/}{Creative Commons Attribution 4.0 International License}.}
    \label{fig:phase-transformation-slender}
\end{figure}

\noindent Fig.~\ref{fig:phase-transformation-slender} shows the phase transformation of a slender tape-like salicylideneamine crystal of dimensions {$1440 \times 360 \times 36 $ $\mathrm{\mu m^3}$} under an incident UV light. This molecular crystal is fixed with $\mathbf{u} = 0$ on (0$\bar{1}$0) face, and is uniformly illuminated under UV light (365nm, 60mWcm$^{-2}$) on the $(001)$ face, see {Fig.~\ref{fig:LightDecay-PartGeometry}(b)}. The incident UV radiation first drives the enol-to-keto isomerization reaction (represented by the order parameter $c$), see {Fig.~\ref{fig:phase-transformation-slender}(b)}, which in turn triggers the $\beta$-to-$\gamma$ phase transformation at room temperature (represented by the structural order parameter $\eta$), see {Fig.~\ref{fig:phase-transformation-slender}(a)}. 

\begin{figure}
    \centering
    \includegraphics[width=0.6\textwidth]{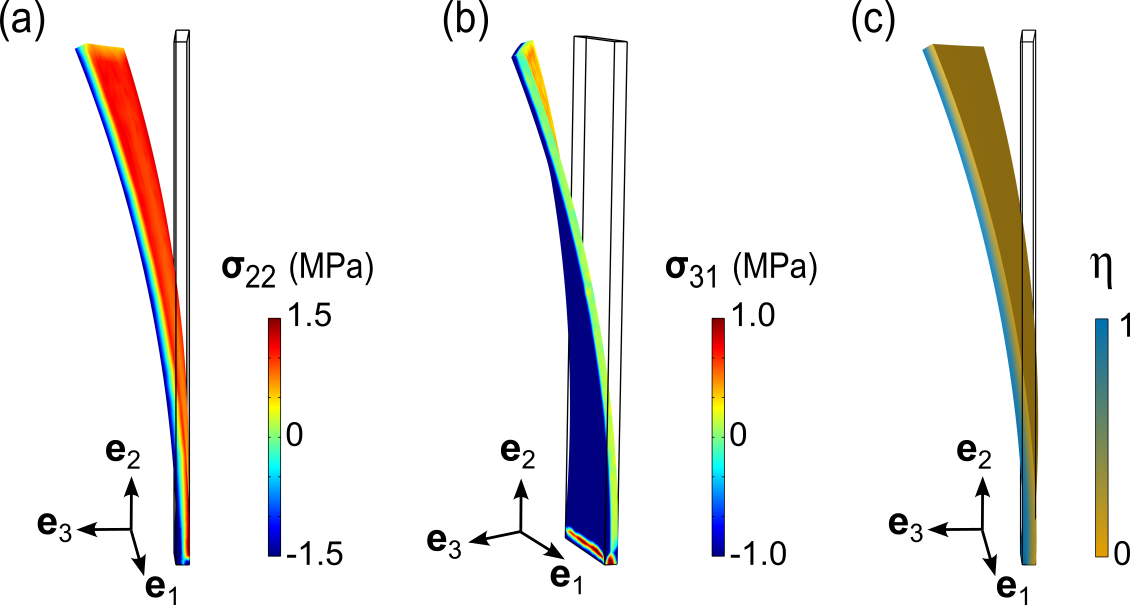}
    \caption{(a-b) Axial and shear stress components in a partially transformed salicylideneamine crystal at $\langle \eta \rangle = 0.5$. (a) In-plane stress ($\sigma_{22}$) arise from the lattice misfit between the keto molecules on the surface and the underlying molecular crystal. (b) Shear stress ($\sigma_{31}$) result from the lattice misfit between the $\beta$ and $\gamma$ phases at the phase boundary. (c) These stresses at $\langle \eta \rangle = 0.5$ collectively generate bending and twisting deformations in the molecular crystal.}
    \label{fig:stress Thin Plate}
\end{figure}

\vspace{2mm}
\noindent Fig.~\ref{fig:phase-transformation-slender}(a,c) shows the different stages of photo-induced phase transformation in the salicylideneamine crystal. Under a UV light stimulus, a thin layer of keto molecules ($c = 1$) nucleate on the surface of the molecular crystal, see Fig.~\ref{fig:phase-transformation-slender}(b). These molecules form as a result of the barrierless photoisomerization reaction in the photoexcited state, see Fig.~\ref{fig:MultDecomp-FreeEnergy}(c), and are stable under the UV light stimulus. The difference in enol and keto molecular conformation generates in-plane stresses of $\sigma_{22} = {1.5\times10^{6}}\, \mathrm{Nm^{-2}}$ on the surface of the molecular crystal and bends the crystal toward the light stimulus (see Fig.~\ref{fig:stress Thin Plate}(a)). These in-plane stresses induce the $\beta \to \gamma$ phase transformation in Fig.~\ref{fig:phase-transformation-slender}(a).

\vspace{2mm}
\noindent Fig.~\ref{fig:phase-transformation-slender}(a) shows the nucleation and growth of the $\gamma-$phase through the depth of the molecular crystal. A planar phase boundary with $\hat{\mathbf{n}} = (0,0,1)$ separates the two phase microstructure and this coexistence of the $\beta$ and $\gamma$ phases is associated with the twisting deformation. 
The deformation gradient corresponding to the $\beta \to \gamma$ transformation contains non-diagonal terms (see {Eq.~\ref{FSalicyli}}) that generates lattice misfit at the phase boundary (and at the fixed end of the molecular crystal). 
This lattice misfit generates interfacial shear stresses that twist the slender molecular crystal, see {Fig.~\ref{fig:stress Thin Plate}(b)}.
The twisting deformation is transient, i.e., observed only in the two-phase microstructural state of the molecular crystal, and disappears on complete phase transformation, {see {Fig.~\ref{fig:phase-transformation-slender}(a)}}\cite{taniguchi2019photo}. 
With continued illumination, the keto layer persists on the {(001)} surface of the molecular crystal ($\sim 5\%$ depth) and causes the large sideways bending of the molecular crystal, {see {Fig.~\ref{fig:phase-transformation-slender}(b)}}. 
This bending and twisting deformation of the molecular crystal is consistent with the experimental observations by Taniguchi et al. \cite{taniguchi2019photo}.

\begin{figure}
    \centering
    \includegraphics[width=0.5\textwidth]{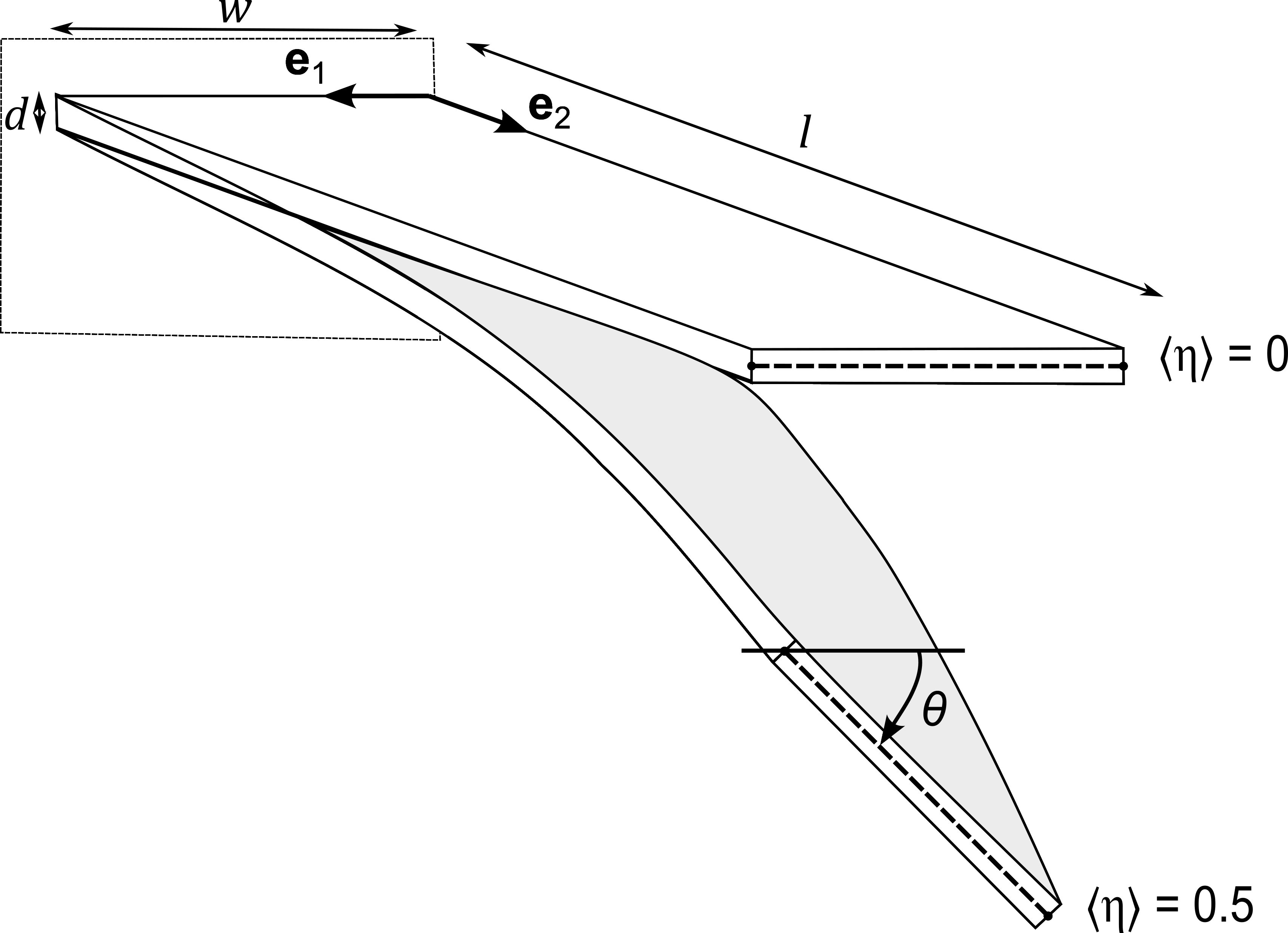}
    \caption{A schematic illustration depicting the twisting deformation of a slender prismatic beam, fixed at one end and subjected to a torque at the other. The twisting induces shear stresses across the beam's cross-section. In our calculations, the solid-to-solid phase transformation in the molecular crystal generates shear stresses that contribute to the twisting deformation.}
    \label{fig:Twist Schematic}
\end{figure}

\vspace{2mm}
\noindent The twisting of the molecular crystal is a consequence of the internal shear stresses and the slender beam-like geometry. For example, at halfway through the phase transformation, i.e., $\langle \eta \rangle = 0.5$, the molecular crystal twists by an angle of $\theta = 19.41^\circ$ at its free end, {see Fig.~\ref{fig:Twist Schematic}}. This twisting deformation is analogous to the torsion of a prismatic beam. That is, a beam with rectangular cross-section $w \times d$ twists by an angle $\theta$ under an applied torque $\mathrm{T}$ as $\theta = \frac{\mathrm{T} l}{\mathrm{k}_1 G wd^3}$ \cite{ugural2003advancedAnalytTwistbook}. This twist angle is also related to the maximum shear stresses $\tau_{\mathrm{max}}$ by $\theta = \frac{\tau_{\mathrm{max}} \mathrm{k}_2 l}{\mathrm{k}_1 G d}$. Here, $G$ is the shear modulus of the molecular crystal and $\mathrm{k}_1, \mathrm{k}_2$ are non-dimensional numbers related to geometry of the prismatic beam (see Appendix \ref{appendix:torsion}) \cite{ugural2003advancedAnalytTwistbook}. By substituting the molecular crystal dimensions and shear stresses generated during phase transformation we obtain $\theta = 22.91^\circ$. This compares favorably with our computational result of $\theta = 19.41^\circ$. We attribute the difference in twist angles between our analytical and computational results to the combined bending and twisting deformation in our molecular crystal and the presence of a two-phase microstructural state at $\langle \eta \rangle = 0.5$. 

\vspace{2mm}
\noindent On turning off the light stimulus, i.e., $\mathcal{I} = 0$, both the photoisomerized keto molecules and the $\gamma-$phase are energetically unstable and the molecular crystal transforms to its initial $\beta-$phase. In absence of light stimulus the keto molecules slowly back-isomerize to the enol form and trigger the $\gamma \to \beta$ reverse transformation. This reverse $\gamma \to \beta$ transformation is much slower, compared to the forward $\beta \to \gamma$ transformation, owing to the small mobility associated with the energy minimization process ($\mathcal{M}_2 \ll \mathcal{M}_1$) and the absence of photoreaction driving forces. The microstructures, see {Fig.~\ref{fig:phase-transformation-slender}(a)}, retrace their two-phase coexistence and deform sequentially by bending and twisting of the molecular crystal. Eventually, the molecular crystal fully recovers its initial configuration and returns to the energy minimizing $\beta-$phase. 

\vspace{2mm}
\noindent In line with experimental literature \cite{taniguchi2019photo}, our computations predict a two-phase microstructure to form in the molecular crystal during phase transformation. Our continuum model based on energy minimization across a multi-well energy landscape (that accounts for a local minima corresponding to the photoexcited keto-molecules) and without a priori assumptions, predicts the bending and twisting deformation in the salicylideneamine crystal. Using the lattice deformation gradient (corresponding to the $\beta \to \gamma$ transformation) and thermodynamic energy barriers as inputs (and without prescribing the interface boundary conditions explicitly), our model predicts the bending and twisting deformation observed in the salicylideneamine crystal. These results are consistent with the photomechanical deformation observed by Taniguchi et al. \cite{taniguchi2019photo} and provides additional insights into the stresses accompanying phase transformation. We next apply our modeling framework to investigate whether and how factors such as, particle geometry and light intensity, affect phase transformation and deformation pathways. 

\subsection{Particle Geometry}
\begin{figure}
    \centering
    \includegraphics[width=\textwidth]{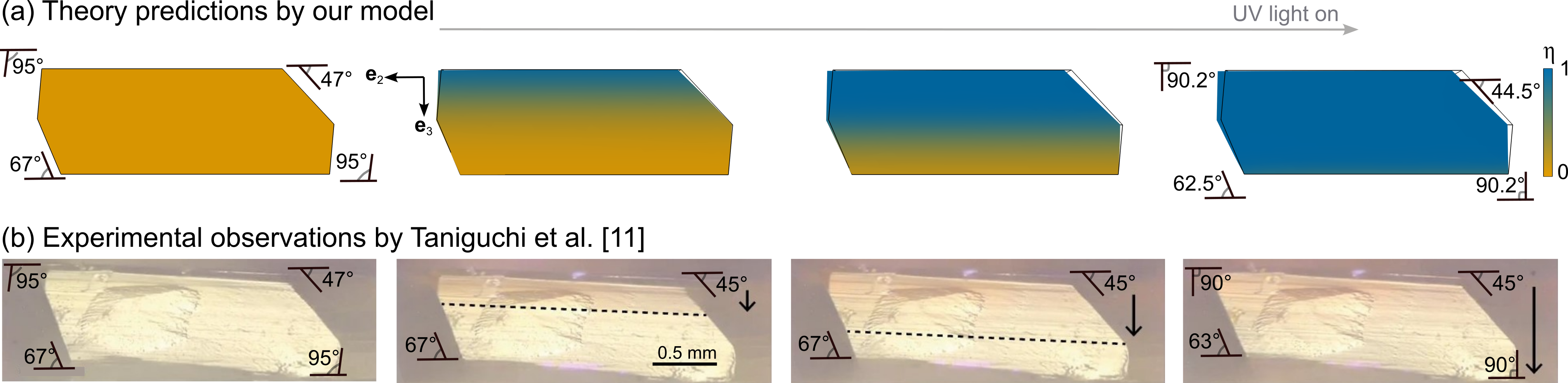}
    \caption{(a-b) A salicylideneamine crystal with a thick platelet-like geometry undergoes shearing deformation during the phase change. Our theoretical predictions of this shearing deformation, along with the accompanying changes in particle geometry, are consistent with the experimental measurements reported by Taniguchi et al. \cite{taniguchi2019photo}; reprinted from Ref.~\cite{taniguchi2019photo} under a \href{https://creativecommons.org/licenses/by/4.0/}{Creative Commons Attribution 4.0 International License}. These calculations validate the use of Cauchy-Born rule in predicting the macroscopic deformation of the salicylideneamine crystal.}
    \label{fig:phase-transformation-thick}
\end{figure}

{Fig.~\ref{fig:phase-transformation-thick}(a-b)} shows the photo-triggered macroscopic deformation of a thick platelet-like salicylideneamine crystal. Following the experimental specimens in Ref.~\cite{taniguchi2019photo}, we model a molecular crystal of dimensions {$2010 \times 1235 \times 737 $ $\mathrm{\mu m^3}$} with facets along (0$\bar{1}$$\bar{1}$) and (012) directions, see Fig.~\ref{fig:LightDecay-PartGeometry}(b). As in {Fig.~\ref{fig:LightDecay-PartGeometry}}(b), the bottom face (00$\bar{1}$) of the crystal is fixed with $\mathbf{u}=0$, and the top face (001) of the crystal is uniformly illuminated under UV light {(365nm, 80mWcm$^{-2}$)}. Except for the particle geometry, all other modeling parameters are the same as for the computations shown in {Fig.~\ref{fig:phase-transformation-slender}(a)}.

\vspace{2mm}
\noindent {Fig.~\ref{fig:phase-transformation-thick}(a)} shows a two-phase microstructure in the thick molecular crystal undergoing a photo-triggered $\beta \to \gamma$ phase transformation. The phase boundary, separating the $\beta$ and $\gamma$ phases, is planar and propagates approximately parallel to the (001) plane. This orientation of the phase boundary is consistent with our analytical calculations of the coherent interface in {section~\ref{sec:Cauchy-Born}} and with the experimental measurements reported by Taniguchi et al. \cite{taniguchi2019photo}, {see Fig.~\ref{fig:phase-transformation-thick}(a-b)}. 

\begin{figure}
    \centering
    \includegraphics[width=\textwidth]{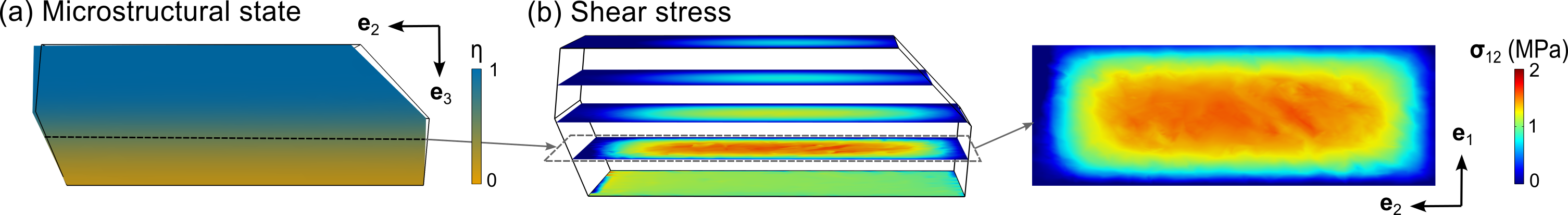}
    \caption{(a-b) The microstructural state and shear stress distribution ($\sigma_{12}$) in a thick platelet-like crystal during phase transformation {$\langle\eta\rangle=0.75$}. The shear stresses are maximum at the phase boundary and primarily arise from the lattice mismatch between the $\beta$ and $\gamma$ phases.}
    \label{fig:stress Thick crystal}
\end{figure}

\vspace{2mm}
\noindent The $\beta \to \gamma$ phase transformation in {Fig.~\ref{fig:phase-transformation-thick}} is accompanied by an affine shearing of the thick crystal. As before, the lattice mismatch between the $\gamma$ and $\beta$ phases generates a shear stress $\sigma_{12} = {2\times10^{6}}\, \mathrm{Nm^{-2}}$ on the (001) plane, see {Fig.~\ref{fig:stress Thick crystal}(b)}. This interplay between the shear stresses, arising from internal driving forces, and the particle geometry of the molecular crystal manifests as a shearing deformation. Fig.~\ref{fig:phase-transformation-thick} presents the deformed configurations of the crystal that are geometrically consistent with the experimental observations \cite{taniguchi2019photo}. For example, the angle between the crystal facets of the platelet particle changes as a function of the phase transformation, see Fig.~\ref{fig:phase-transformation-thick}(a). This macroscopic deformation aligns closely with the experimentally reported values, as shown in Fig.~\ref{fig:phase-transformation-thick}(b). We note that in thicker platelet-like molecular crystals, the presence of surficial keto-molecules does not significantly alter the macroscopic deformation of the molecular crystal. 

\begin{figure}
    \centering
    \includegraphics[width=0.80\textwidth]{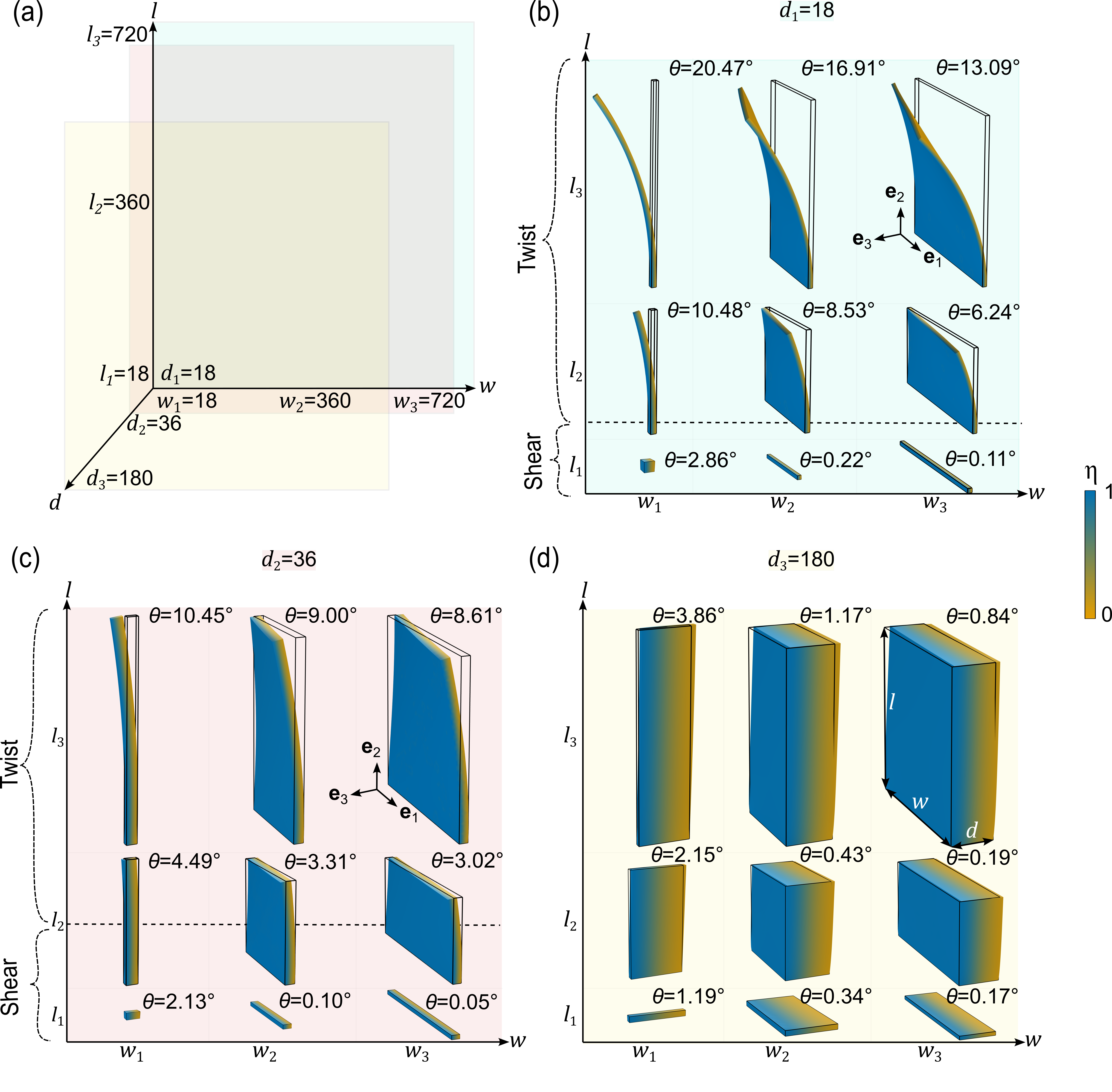}
    \caption{(a-d) Phase transformation microstructures in salicylideneamine crystals with varying particle geometries $l \times w \times d$. The internal shear stresses in these crystals at $\langle \eta \rangle = 0.5$ generate a range of deformations, from bending and twisting in slender geometries to shearing in thicker, plate-like geometries. In all cases, the molecular crystal is uniformly illuminated on the (001) face and fixed on the $\mathbf{e}_1-\mathbf{e}_3$ plane. Subfigures (b-d) correspond to molecular crystals with thickness of $d = 18, 36,$ and $180 \, \mu \mathrm{m}$, respectively.}
    \label{fig:Particle Geometry}
\end{figure}

\vspace{2mm}
\noindent We next systematically vary the particle geometries of the molecular crystal in the range $18 \mathrm{\mu m} \le {d} \le 180\mathrm{\mu m}$, $18\mathrm{\mu m} \le {w} \le 720\mathrm{\mu m}$, and $18\mathrm{\mu m} \le {l} \le 720\mathrm{\mu m}$ (see Fig.~\ref{fig:Particle Geometry}) and compute the $\beta \to \gamma$ phase transformation under a light stimulus ($\lambda$ = 365nm, $\mathcal{I}$ = 80mWcm$^{-2}$). In all cases, we fix the bottom face of the molecular crystal and illuminate the {(001)} face. Fig.~\ref{fig:Particle Geometry} shows the macroscopic deformation of the salicylideneamine crystals, at $\langle \eta \rangle = 0.5$, as a function of particle geometry. In each computation, the lattice mismatch between the $\gamma$ and $\beta$ phases generates a shear stresses in the range {$1\times10^6$ Nm$^{-2}$ to $6\times10^6$ Nm$^{-2}$} on the (001) plane. These stresses generate a significant twisting deformation in slender crystals (with ${l}/{d} \gg 10$) and a shearing deformation in thicker crystals (with $\sim l/d < 10$). The handedness of the twist is clock-wise as shown in Fig.~\ref{fig:Twist Schematic}, and is a consequence of a single lattice variant generated during the triclinic-to-monoclinic transformation of the multi-lattices, see {section~\ref{sec:multi-lattice}.}

\subsection{Incident Light}\label{sec:IncidentLight}
\begin{figure}
    \centering
    \includegraphics[width=\textwidth]{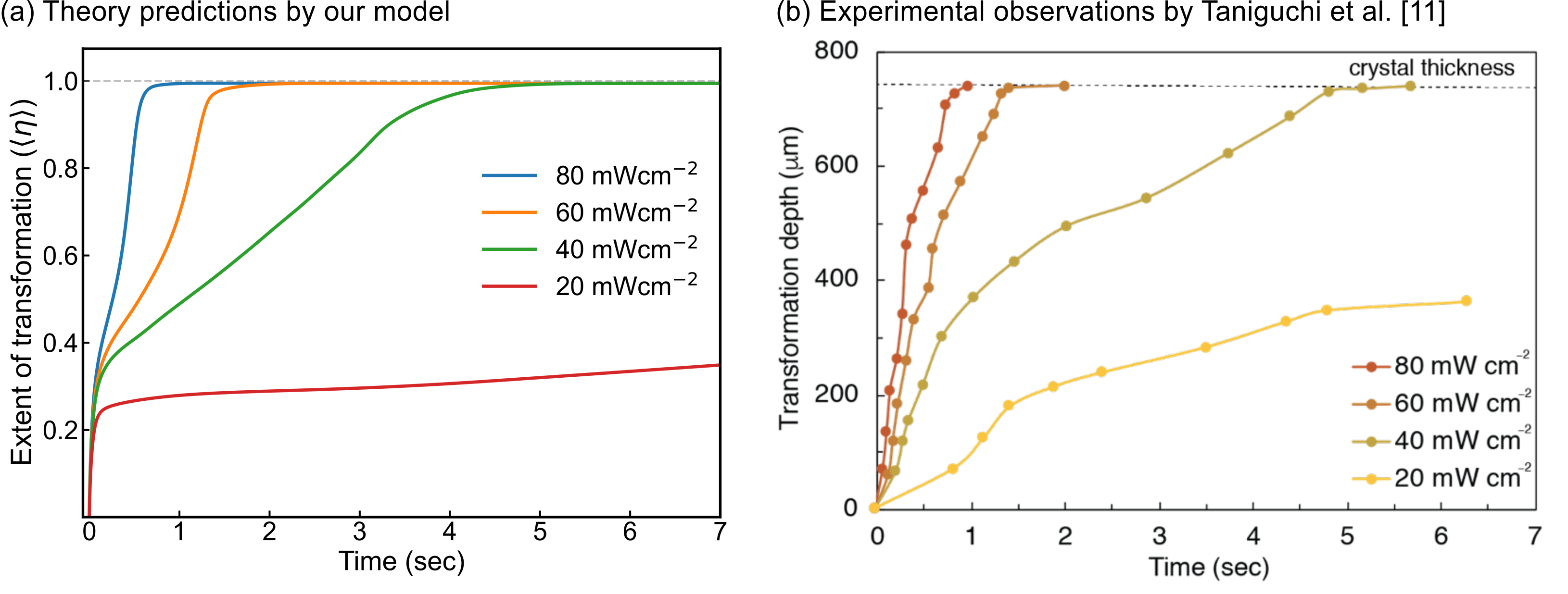}
    \caption{(a) Temporal profiles showing the extent of phase transformation in a thick, platelet-like salicylideneamine crystal ($2010 \times 1235 \times 737 $ $\mathrm{\mu m^3}$) under different intensities of the incident UV light. (b) The extent of phase transformation along the thickness of the salicylideneamine crystal over time, under various UV light intensities as observed by Taniguchi et al. \cite{taniguchi2019photo}.  Reprinted from Ref.~\cite{taniguchi2019photo} under a \href{https://creativecommons.org/licenses/by/4.0/}{Creative Commons Attribution 4.0 International License}. Our theory-predicted temporal profiles are consistent with the linearized phase transformation kinetics reported in the experiments \cite{taniguchi2019photo}. }
    \label{fig:Time-Eta}
\end{figure}

We next investigate the effect of incident light intensity on phase transformation in molecular crystals, 
see Fig.~\ref{fig:Time-Eta}. A thick platelet-like salicylideneamine crystal of dimensions {$2010 \times 1235 \times 737 $ $\mathrm{\mu m^3}$} is illuminated with UV light ($\lambda = 365~\mathrm{nm}$) as shown in Fig.~\ref{fig:LightDecay-PartGeometry}(b). Please recall that the incident light decays through the depth of the molecular crystal in accordance with the Beer-Lambert law as shown in Fig.~\ref{fig:LightDecay-PartGeometry}(a).

\vspace{2mm}
\noindent Fig.~\ref{fig:Time-Eta}(a) shows the rate of phase transformation in a representative salicylideneamine crystal under light intensities $20 \le \mathcal{I}_0 \le 80\mathrm{mWcm}^{-2}$. In all cases, we note that illuminating the salicylideneamine crystal in its reference $\beta-$phase nucleates a $\gamma-$phase that grows rapidly in the initial stage and saturates at finite values of $\langle\eta\rangle$, see Fig.~\ref{fig:Time-Eta}(a). For light intensities $\mathcal{I}_0 \leq 20\mathrm{mWcm}^{-2}$, prolonged illumination of the molecular crystal does not significantly grow the $\gamma-$phase and the extent of transformation plateaus before completion. This incomplete transformation indicates that there is a threshold of the incident light intensity that is necessary to complete the $\beta \to \gamma$ phase transformation. For light intensities {$\mathcal{I}_0 \geq 40\mathrm{mWcm}^{-2}$}, the internal driving forces are sufficient to overcome the energy barrier between the $\beta \to \gamma$ phases leading to complete phase transformation, see Fig.~\ref{fig:Time-Eta}. 

\vspace{2mm}
\noindent Our modeling predictions for the temporal profile of the extent of transformation, as shown in Fig.~\ref{fig:Time-Eta}(a), are consistent with several features observed in the experimental measurements by Taniguchi et al. shown in Fig.~\ref{fig:Time-Eta}(b)\cite{taniguchi2019photo}. For example, our model predicts an exponential growth of the $\gamma-$phase during the initial stage of phase transformation and an eventual saturation with $\langle \eta \rangle \approx50\%$ for incident light intensities $\mathcal{I}_0 \leq 20\mathrm{mWcm}^{-2}$. At higher light intensities $40 \le \mathcal{I}_0 \leq 80\mathrm{mWcm}^{-2}$, our model predicts a rapid $\beta \to \gamma$ phase transformation in the molecular crystal. The phase transformation kinetics scales linearly with increasing light intensities and matches the experimentally observed transformation completion times reported in Taniguchi et al. \cite{taniguchi2019photo}, see Figs~\ref{fig:Time-Eta}(a-b). 

\vspace{2mm}
\noindent The quantitative differences between the theoretical predictions and experimental measurements of phase transformation kinetics in Figs.~\ref{fig:Time-Eta}(a-b) can be attributed to the following factors: The first is related to the order of photoisomerization kinetics and the $\beta-\gamma$ transformation kinetics that the molecular crystal follows under different intensities of incident light. For example, previous works show that the phase transformation kinetics in molecular crystals is highly nonlinear and that the order of reaction kinetics is not a constant within a single phase transformation cycle \cite{reactionKineticsBardeen}. A second factor, could be related to the photothermal effects on phase transformation kinetics. Taniguchi et al. \cite{taniguchi2019photo} show that the temperature of the illuminated surface in salicylideneamine changes over $10^\circ${C} during the $\beta\to\gamma$ phase change, which is likely to affect the mobility of the phase boundary.\footnote{It is important to note that, despite the variations in temperature, the salicylideneamine crystal consistently remained below its phase transition temperature of 39.6$^\circ${C} \cite{taniguchi2019photo}. This observation reinforces the conclusion that the $\beta \to \gamma$ phase transformation in the salicylideneamine crystal was triggered by a light stimulus rather than thermal effects \cite{taniguchi2019photo}.} These factors would affect phase transformation kinetics and should be considered in engineering photomechanical materials.

\newpage
\section{Discussion}
We developed a continuum model to predict the photo-triggered deformations in molecular crystals. A distinguishing feature of our framework is that we construct a multi-well energy landscape---by accounting for local minima corresponding to the photoexcited states---and use the lattice-continuum link (Cauchy-Born rule) to predict the solid-to-solid phase transformations in molecular crystals.
By minimizing the total free energy of the system across a multi-well energy landscape, combined with photoreaction kinetics, we compute the subtle interplay between lattice transformation pathways, potential energy minimization, and microstructural patterns in governing the macroscopic deformation of salicylideneamine crystals. The theoretical framework is general and can be adapted to describe photo-triggered phase changes in other molecular crystals. Below, we discuss a few shortcomings of our model and outline the relevance of our findings on the micromechanisms governing photomechanical deformation.

\vspace{2mm}
\noindent In our model, we assume the Cauchy-Born rule in mapping the deformation gradient of constituent Bravais lattices to the continuum deformation of the molecular crystal. A consequence of this assumption is that the shift vectors, which describe the shuffling in multi-lattices, are not linked with the macroscopic deformation of the crystal \cite{james1987shiftvector, bhattacharya2003microstructure, bhattacharya1997relaxation}. Under quasi-static equilibrium conditions, this is a reasonable approximation in which shifts adjust themselves locally in order to minimize the total energy for any given deformation of the system. For the salicylideneamine crystal \cite{taniguchi2019photo}, the shift vector can be formulated as an internal variable that is a function of the structural order parameter $\eta$ {(see Appendix~\ref{Appendix:Multi-lattice})}. However, in other molecular crystals that transform under non-equilibrium conditions, the shift vectors could have a profound effect on the macroscopic deformation of a system. It is therefore important to formulate an energy landscape that is not only a function of $\mathbf{F}$ and $\eta$, but also a function of the shifts $\mathbf{p}$, to accurately determine the macroscopic deformation and phase transformation pathways in molecular crystals. 

\vspace{2mm}
\noindent Our continuum model does not consider the changing illumination angles resulting from the finite deformations of a molecular crystal. For example, the bending and twisting of the salicylideneamine crystal is accompanied by large rigid body rotations, which, in turn, alters the incident angle of light on the crystal surface. This interplay did not significantly affect the macroscopic response of the salicylideneamine crystal (see Refs.~\cite{taniguchi2019photo, taniguchi2022superelasticity}), however, would play an important role in predicting the deformation of other crystals with a light-interactive molecular dipole \cite{bardeenIncidentangle}. By accounting for the change in relative orientation between the crystal and light stimulus as an internal variable, the continuum model could be extended to predict the nucleation and growth of heterogeneous microstructures in molecular crystals and its effect on the macroscopic deformation behavior \cite{bardeenIncidentangle, maghsoodiLCEs}.

\vspace{2mm}
\noindent With these reservations in mind, we next highlight the significance of our findings on photo-triggered deformations in molecular crystals. Our continuum model provides a framework for investigating the interplay between thermodynamic (and elastic) energy minimization and photoexcited states on a multi-well energy landscape. Specifically, by accounting for a local minima corresponding to the keto molecules, we capture the interplay between the isomerization reaction and the structural phase transformation pathways in a molecular crystal. This coupling was important to overcome the thermodynamic energy barriers governing the $\beta \to \gamma$ phase transformation in the salicylideneamine crystal. By integrating these effects in our model, we present a comprehensive understanding of the micromechanisms that generate the bending and twisting deformations in slender salicylideneamine crystals.

\section{Conclusion}
We demonstrate that the macroscopic behavior of molecular crystals in response to a light stimulus is closely linked to the structural transformation of lattices at the atomic scale. Our continuum model, based on the Cauchy-Born rule and photoreaction theory, uses lattice geometries as the primary input to predict experimentally consistent bending, twisting, and shearing deformation in salicylideneamine molecular crystal (a representative material). Central to our modeling framework is a multi-well energy landscape that accounts for photoexcited states (e.g., keto-molecules), thermodynamic energy barriers, and finite elastic
deformations that contribute to the total energy of the system. By integrating these effects and minimizing the energy across a multi-well landscape, our model predicts the bending and twisting deformation in salicylideneamine crystals without any a priori conditions. Our findings on the microstructural evolution pathways, investigated across varying particle geometries and light intensities, are consistent with experimental measurements and identify geometric regimes that generate twisting and shearing deformations. Our model provides a theoretical framework for investigating the micromechanisms governing kinematic behaviors in molecular crystals and shows how a continuum model could serve as a design tool to advance the design of molecular crystals with controllable and reversible deformation.

\section{Acknowledgment}

D. Tiwari and A. Renuka Balakrishna gratefully acknowledge the research funding by the Air Force Fiscal Year 2023 Young Investigator Research Program, United States, under Grant No. FOA-AFRL-AFOSR-2022-0005 (DT, ARB). This work was partially supported by the National Science
Foundation through the Materials Research Science and Engineering Center (MRSEC) at UC Santa Barbara: NSF DMR–2308708 (IRG-X/Seed). The authors thank the Center for Scientific Computing at University of California, Santa Barbara (MRSEC; NSF DMR 2308708) for providing computational resources that contributed to the results reported in this paper.

\newpage

\appendix

\section{Free Energy Function}\label{Appendix:EnergyNorm-ParameterTable}
In this section we non-dimensionalize the free energy function used in our micromechanical model and list the material constants corresponding to the salicylideneamine crystal. 

\vspace{2mm}
\noindent The free energy is a function of the photoisomerization order parameter $c$, a structural order parameter $\eta$, and Green-Lagrange strain tensors $\mathbf{E}, \bar{\mathbf{E}}$. We divide the free energy function in Eq.~\ref{Eq:freeEnergy} by the thermodynamic energy coefficient, $A_1$:

\begin{align}
    \frac{\Psi}{A_1} &= \int_\Omega \bigl\{ {\frac{1}{2}\nabla c\cdot\left(\frac{\mathbf{K}_1}{A_1}\right)\nabla c + \frac{1}{2}\nabla\eta\cdot\left(\frac{\mathbf{K}_2}{A_1}\right)\nabla\eta} + { [\eta^2 (1-\eta)^2] + \left(\frac{A_2}{A_1}\right) \left[\frac{\eta^2}{2}-\frac{\eta^3}{3}\right]} \nonumber\\
    & + \left(\frac{A_3}{A_1}\right)[c^2(1-c)^2] + \left(\frac{A_4}{A_1}\right)(1-\eta^2) c^2 + \frac{1}{2}J_0[\bar{\mathbf{E}}:\left(\frac{\mathbb{C}}{A_1}\right)\bar{\mathbf{E}}] - \left(\frac{1}{A_1}\right)(\mathbf{\sigma}_0(c):\mathbf{E}) \bigr\}\mathrm{d}\mathbf{x}.
\end{align}

\noindent The governing equations that describe photoisomerization kinetics and $\beta \to \gamma$ phase transformation kinetics are given by Eq.~\ref{Eq:isomerizationKinetics} and Eq.~\ref{Eq:kinetics}, respectively. We normalize these kinetic equations as follows: 

\begin{align}
    \left(\frac{1}{\mathcal{M}_1 A_1}\right) \frac{\partial c}{\partial t}  = - \left[ \frac{\partial ({\Psi}/{A_1})}{\partial c} -  { \left(\frac{\mathbf{K}_1}{A_1}\right) \nabla^2 c} \right]  + \left(\frac{1}{\mathcal{M}_1 A_1}\right)\Upsilon \mathcal{I} f
    \label{eq:normalizedIsomerizationkinetics}
\end{align}

\begin{align}
    \left(\frac{1}{\mathcal{M}_2 A_1}\right) \frac{\partial \eta}{\partial t}  = - \left[ \frac{\partial ({\Psi}/{A_1})}{\partial \eta} -  { \left(\frac{\mathbf{K}_2}{A_1}\right) \nabla^2 \eta} \right]  + \left(\frac{1}{\mathcal{M}_2 A_1}\right){\chi (1-\eta)}
    \label{eq:normalizedPhasekinetics}
\end{align}

\noindent Please note that in solving for mechanical equilibrium in Eq.~\ref{Eq:elasticityTensor}, we use the non-dimensional form of the stiffness tensor, $\left(\frac{\mathbb{C}}{A_1}\right)$. 

\vspace{2mm}
\noindent The micromechanical model developed in this work is calibrated for salicylideneamine molecular  crystal at room temperature. We derive a majority of the material constants from the systematic experiments conducted by Taniguchi et al. \cite{taniguchi2019photo} and summarize them in Table~\ref{Table:model-parameters}.

\begin{table}[h!]
\centering
\begin{tabular}{lcc}
\hline
\textbf{Physical quantities} & \textbf{Notation} & \textbf{Value} \\
\hline
Thermodynamic energy constants    & ${A_1} = {\Psi_0}$ & $2.5 \times {10}^6  \mathrm{~J} / \mathrm{m}^3 $ \cite{taniguchi2019photo}\\
($\beta\to\gamma$ phase change) & ${A_2}/{A_1}$ & 1.2 \\
&& \\
Photoisomerization constants & ${A_3}/{A_1}$ & 10 \\
(enol to keto) & ${A_4}/{A_1}$ & 1 \\
&& \\
Gradient energy coefficient & $\sqrt{\frac{\mathbf{K}}{A_1}}$ & $10^{-6}\mathrm{m}$ \\
Absorption coefficient& $\mu$ & $0.4 {(\mu \mathrm{m})^{-1}}$  \\
&&\\
Shear modulus & $\mathrm{G}$ & $1.0 \times {10}^8  \mathrm{~Pa} $ \cite{taniguchi2022superelasticity} \\
Poisson's ratio & $\nu$ & $0.2 $ \cite{taniguchi2022superelasticity} \\
\hline
\end{tabular}
\caption{Physical quantities and material constants of the salicylideneamine crystal used in our model.}
\label{Table:model-parameters}
\end{table}

\section{Governing Equations}\label{Appendix:Weakform}
The weak form of the kinetic equations and mechanical equilibrium condition solved in our model is presented below. We derive these weak forms by integrating Eq.~\ref{eq:normalizedIsomerizationkinetics} over the reference volume $\Omega$ and using a test function $\phi$ to satisfy the appropriate smoothness conditions. For photoisomerization reaction kinetics, we have:

\begin{align}
\int_{\Omega}  \left(\frac{1}{\mathcal{M}_1 A_1}\right) \left(\frac{\partial c}{\partial t}\right) \phi \, d\mathbf{x} & = -\int_{\Omega} \left \{ \left( \frac{\partial ({\Psi}/{A_1})}{\partial c} \right) \phi + {\left(\frac{\mathbf{K}_1}{A_1}\right)} {\nabla} (c)  {\nabla} ( \phi ) - \left(\frac{1}{\mathcal{M}_1 A_1}\right)\Upsilon \mathcal{I} f \phi \, \right\} \mathrm{d}\mathbf{x} \nonumber \\
& \quad + \int_{S_0}  \left(\frac{\mathbf{K}_1}{A_1}\right) \left({\nabla} (c) . \hat{\mathbf{n}}\right) ) \phi \, \mathrm{d}{S}.
\label{Eq:IsoweakForm}
\end{align}
\noindent Similarly, the weak form for the phase transformation kinetics (see Eq~\ref{eq:normalizedPhasekinetics}) is derived as:
\begin{align}
\int_{\Omega}  \left(\frac{1}{\mathcal{M}_2 A_1}\right) \left(\frac{\partial \eta}{\partial t}\right) \phi \, \mathrm{d}\mathbf{x} & = -\int_{\Omega} \left \{ \left( \frac{\partial ({\Psi}/{A_1})}{\partial \eta} \right) \phi + {\left(\frac{\mathbf{K}_2}{A_1}\right)} {\nabla} (\eta)  {\nabla} ( \phi ) - \left(\frac{1}{\mathcal{M}_2 A_1}\right)\chi (1-\eta) \phi \, \right\} \mathrm{d}\mathbf{x} \nonumber \\
& \quad + \int_{S_0}  \left(\frac{\mathbf{K}_2}{A_1}\right) \left({\nabla} (\eta) . \hat{\mathbf{n}}\right) ) \phi \, \mathrm{d}{S}.
\label{Eq:PhaseweakForm}
\end{align}

\noindent The weak for mechanical equilibrium (see Eq.~\ref{eq:mechanical-equilibrium}) is:

\begin{equation}
    \int_{\Omega} \mathbf{P} \nabla\phi \mathrm{d}{x} - \int_{S}   (\mathbf{P} \hat{\mathbf{n}})\phi   \mathrm{d}{S}  = 0.
\end{equation}

\section{Multi-Lattices}\label{Appendix:Multi-lattice}

Fig.\ref{fig:crystalline-order} shows the long-range periodic ordering of molcules in the salicylideneamine crystal with the $\beta$ and $\gamma$ phases corresponding to the reference and transformed configurations, respectively. Please note that the salicylideneamine crystal structure is a multi-lattice with two congruent Bravais lattices that are linearly offset by a shift vector, as discussed in section~\ref{sec:multi-lattice}. The unit cell parameters and shift vectors in the reference and transformed phases are listed in Table~\ref{Table:lattice_parameters}~\cite{taniguchi2019photo}. 

\vspace{2mm}
\noindent Using the unit cell parameters, we derive the lattice vectors in reference $\beta-$phase ($\mathbf{e}_1$, $\mathbf{e}_2$, $\mathbf{e}_3$) as:

\[
\mathbf{e}_1 = \begin{pmatrix}
6.174 \\
0 \\
0
\end{pmatrix},
\quad
\mathbf{e}_2 = \begin{pmatrix}
0.302 \\
9.907 \\
0
\end{pmatrix},
\quad
\mathbf{e}_3 = \begin{pmatrix}
1.296 \\
1.845 \\
19.471
\end{pmatrix},
\]

\noindent and the lattice vectors in the transformed $\gamma-$phase ($\mathbf{f}_1$, $\mathbf{f}_2$, $\mathbf{f}_3$) as:

\[
\mathbf{f}_1 = \begin{pmatrix}
6.189 \\
0 \\
0
\end{pmatrix},
\quad
\mathbf{f}_2 = \begin{pmatrix}
0 \\
9.871 \\
0
\end{pmatrix},
\quad
\mathbf{f}_3 = \begin{pmatrix}
-2.387 \\
0 \\
19.596
\end{pmatrix}.
\]

\noindent The shift vectors for the reference and transformed phases ($\mathbf{p}$ and $\mathbf{q}$, see Table~\ref{Table:lattice_parameters}) can be expressed in terms of the respective lattice vectors as:
\begin{equation}
    \mathbf{p} = -0.46 \mathbf{e}_1 + 0.44 \mathbf{e}_2 + 0.34 \mathbf{e}_3
\end{equation}

\begin{align}
\mathbf{q} & = -0.46 \mathbf{f}_1 + 0.49 \mathbf{f}_2 + 0.34 \mathbf{f}_3 \nonumber \\
& = -0.46 \mathbf{Fe}_1 + (0.44 + \zeta(\eta)) \mathbf{Fe}_2 + 0.34 \mathbf{Fe}_3  \label{eq:shiftCauchyBorn}
\end{align}
In Eq.~\ref{eq:shiftCauchyBorn}, we introduce $\zeta$ a linear function of the order parameter $\eta$. This function takes values of $\zeta=0$ for the $\beta-$phase ($\eta = 0$) and $\zeta=0.05$ for the $\gamma-$phase ($\eta = 1$). This linear definition enforces the Cauchy-Born rule for the shifts when $\zeta(\eta)$ is fixed for a given microstructural configuration during phase change. Using the different values of $\zeta(\eta)$, we can then describe different equilibrium configurations such as the $\beta$ and $\gamma$ phases on the molecular crystal \cite{james1987shiftvector}. Using this assumption and following the derivation in Ref.~\cite{james1987shiftvector} we eliminate the shift out of the energy formulations for equilibrium calculations i.e., $\tilde{\psi}(\mathbf{F}\mathbf{e}_i,\mathbf{p},\eta,c) \approx \tilde{\psi}(\mathbf{F}\mathbf{e}_i,\tilde{\mathbf{p}}(\mathbf{F}\mathbf{e}_i,\eta,c),\eta,c) \approx\psi(\mathbf{F},\eta,c)$. The free energy therefore only depends on the deformation gradient for a given microstructural state \cite{james1987shiftvector, bhattacharya2003microstructure, bhattacharya1997relaxation}.

\subsection{Point group of salicylideneamine multi-lattice}\label{Appendix:Multi-Lattice Point Group}
As discussed in section~\ref{sec:multi-lattice}, the symmetry-lowering transformation of the Bravais lattices during the $\gamma$ (monoclinic) $\to$ $\beta$ (triclinic) phase transformation theoretically generates two lattice variants, which are related to one another by a $180^\circ$ rotation. However, rotations ($\mathbf{R}$) in the point group of a multi-lattice are a subset of rotations in the point group of constituent Bravais lattice that satisfy additional constraint related to the transformation of the shift vectors as described in Eq.\ref{Eq:pointgroup}. 

\vspace{2mm}
\noindent In Eq.~\ref{Eq:pointgroup}, {$\mathbf{Rf}_i = \mu^j_i\mathbf{f}_j$} describes the set of rotations $\mathbf{R}$ that maps a Bravais lattice $\mathcal{L}(\mathbf{f}_j, \mathbf{o})$ back to itself. This rotation for a monoclinic Bravais lattice is given by: 

\begin{equation}
    \mathbf{R}(\mathbf{f}_2,180^\circ) = \begin{pmatrix}
-1 & 0 & 0 \\
0  & 1 & 0 \\
0  & 0 & -1
\end{pmatrix}. \label{rotation-supplement}
\end{equation}
 
\begin{figure}[H]
    \centering
    \includegraphics[width=\textwidth]{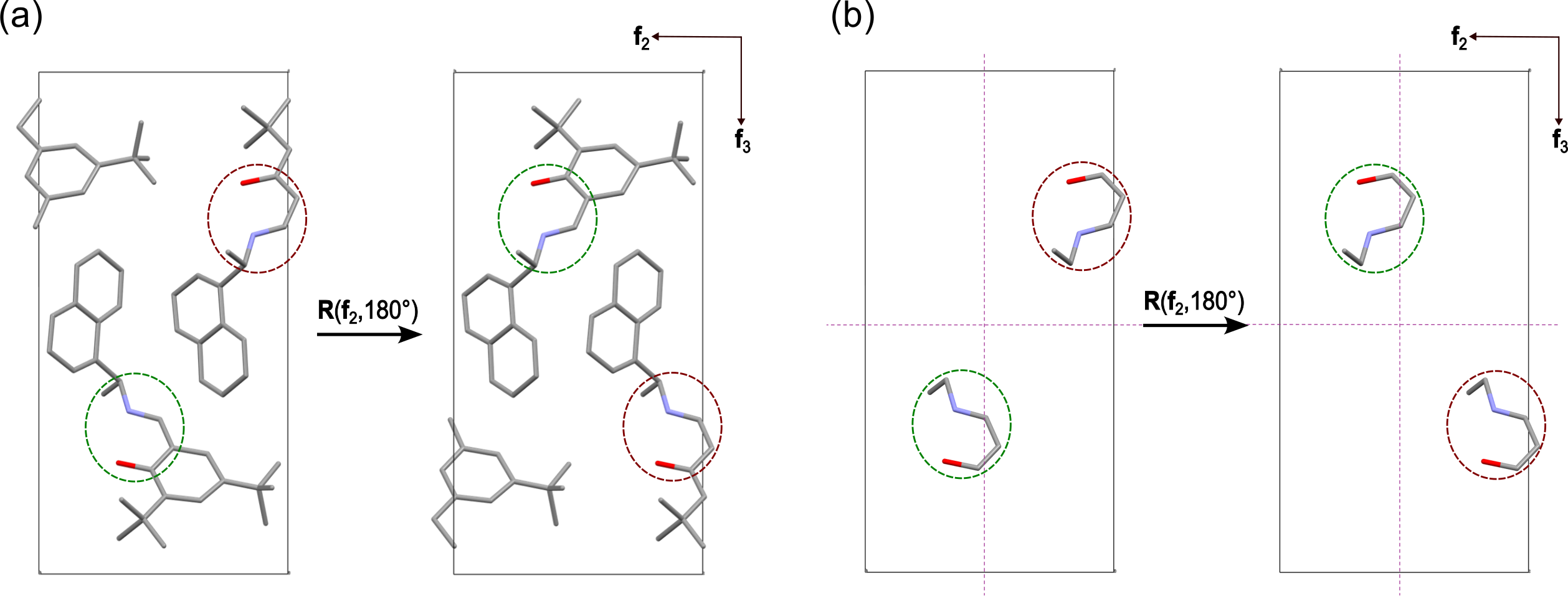}
    \caption{(a) Crystallographic data of the Salicylideneamine crystal {($\gamma-$phase)} with monoclinic symmetry. The two crystallographic representations show the unit cells before and after a $180^\circ$ rotation about the {$\mathbf{f}_2$-axis}. These cells are not superimposable; thus, the rotation $\mathbf{R}(\mathbf{f}_2,180^\circ)$ does not belong to the point group of the monoclinic multi-lattice of the Salicylideneamine crystal in its $\gamma-$phase. Note that the carbon-carbon bonds are shown in gray, the carbon-nitrogen bonds are in blue, the carbon-oxygen bonds are in red, and the hydrogen atoms have been omitted for clarity. (b) Selected molecular parts of the Salicylidenamine crystal are shown to highlight the positional changes that render the two configurations non-superimposable.}
    \label{fig:varaintsExplanation}
\end{figure}

\noindent For the rotation $\mathbf{R}$ in Eq.~\ref{rotation-supplement} to map the multi-lattice $\mathcal{L}(\mathbf{f}_j,\mathbf{q},\mathbf{o})$ back to itself, it must also satisfy the constraint:
\begin{equation}
    \mathbf{R}\mathbf{q}=\nu^i\mathbf{f}_i + \delta\mathbf{q},
    \label{eq:appen-shifttransform1}
\end{equation}
in which, $\nu^i$ are integers and $\delta =1$ \cite{bhattacharya2003microstructure}. However, on substituting $\mathbf{R},\mathbf{f}_i, \mathbf{q}, \, \mathrm{and}\,  \delta =1$ in Eq.~\ref{eq:appen-shifttransform1} and solving for $\nu^i$ gives $\nu^1 = -0.93$, $\nu^2 = 0.00$, $\nu^3 = 0.68$. We can see that the coefficients ($\nu^i$) are non-integers indicating that the rotation $\mathbf{R}$ does not belong to the point group on the salicylideneamine multi-lattice. Consequently, for the salicylideneamine crystal, the only rotation that maps the monoclinic (transformed $\gamma$-phase) multi-lattice to itself is the identity, $\mathbf{R = I}$. Similarly, the triclinic (reference $\beta$-phase) multi-lattice of the molecular crystal is also mapped back to itself solely by $\mathbf{R = I}$. {Moreover, as shown in Fig.~\ref{fig:varaintsExplanation}, the rotation $\mathbf{R}(\mathbf{f}_2,180^\circ)$ alters the molecular configuration, rendering the lattice non-superimposable.} Thus, the number of rotations in the point groups of the triclinic and monoclinic multi-lattices are equal, and consequently the $\beta-\gamma$ symmetry-breaking transformation generates only one variant.

\section{Torsional Parameters}\label{appendix:torsion}
The values of geometric coefficients $\mathrm{k_1}$ and $\mathrm{k_2}$ used in calculating the torsional rigidity of a prismatic beam  {in section~\ref{sec:deformarion-slenderCrystal}} are listed in Table~\ref{tab:torsional-parameters}. These constants correspond to a beam with rectangular cross-section of width $w$ and depth $d$ as shown in Fig.~\ref{fig:Twist Schematic} \cite{ugural2003advancedAnalytTwistbook}:

\begin{table}[ht]
\centering
\begin{tabular}{|c|c|c|c|c|c|c|c|c|c|c|c|}
\hline ${w} / {d}$ & $\mathbf{1 . 0}$ & $\mathbf{1 . 5}$ & $\mathbf{1 . 7 5}$ & $\mathbf{2 . 0}$ & $\mathbf{2 . 5}$ & $\mathbf{3 . 0}$ & $\mathbf{4 . 0}$ & $\mathbf{6 . 0}$ & $\mathbf{8 . 0}$ & $\mathbf{1 0}$ & $\boldsymbol{\infty}$ \\
\hline $\mathrm{k}_{\mathrm{1}}$ & 0.141 & 0.198 & 0.214 & 0.229 & 0.249 & 0.263 & 0.281 & 0.299 & 0.307 & 0.313 & 0.333 \\
\hline $\mathrm{k}_{\mathrm{2}}$ & 0.208 & 0.231 & 0.239 & 0.246 & 0.258 & 0.267 & 0.282 & 0.299 & 0.307 & 0.313 & 0.333 \\
\hline
\end{tabular}
\caption{Torsional parameters for rectangular cross-sections \cite{ugural2003advancedAnalytTwistbook}.}
\label{tab:torsional-parameters}
\end{table}

\newpage

\bibliographystyle{elsarticle-num}

\bibliography{main}
\end{document}